# CMS Paper

# Radiation hardness qualification of $PbWO_4$ scintillation crystals for the CMS Electromagnetic Calorimeter

**The CMS Electromagnetic Calorimeter Group**

P. Adzic[22], N. Almeida[18], D. Andelin[33], I. Anicin[22,a], Z. Antunovic[4], R. Arcidiacono[16], M. W. Arenton[33], E. Auffray[23], S. Argiro[16], A. Askew[40], S. Baccaro[15], S. Baffioni[8], M. Balazs[33], D. Bandurin[35], D. Barney[23], L. M. Barone[15], A. Bartoloni[15], C. Baty[9], S. Beauceron[23], K. W. Bell[29], C. Bernet[8], M. Besancon[7], B. Betev[25], R. Beuselinck[30], C. Biino[16], J. Blaha[9], P. Bloch[23], A. Borisevitch[2], A. Bornheim[38], J. Bourotte[8], R. M. Brown[29], M. Buehler[33], P. Busson[8], B. Camanzi[29], T. Camporesi[30], N. Cartiglia[16], F. Cavallari[15], A. Cecilia[15,b], P. Chang[27], Y. H. Chang[26], C. Charlot[8], E. A. Chen[26], W. T. Chen[26], Z. Chen[25], R. Chipaux[7], B. C. Choudhary[12], R. K. Choudhury[13], D. J. A. Cockerill[29], S. Conetti[33], S. Cooper[36], F. Cossutti[17], B. Cox[33], D. G. Cussans[28], I. Dafinei[15,*], D. R. Da Silva Di Calafiori[25], G. Daskalakis[10], A. David[18], K. Deiters[24], M. Dejardin[7], A. De Benedetti[36], G. Della Ricca[17], D. Del Re[15], D. Denegri[7], P. Depasse[9], J. Descamps[7], M. Diemoz[15], E. Di Marco[15], G. Dissertori[25], M. Dittmar[25], L. Djambazov[25], M. Djordjevic[22], L. Dobrzynski[8], A. Dolgopolov[36], S. Drndarevic[22,a], G. Drobychev[2], D. Dutta[13], M. Dzelalija[4], A. Elliott-Peisert[23], H. El Mamouni[9], I. Evangelou[11], B. Fabbro[7], J. L. Faure[7], J. Fay[9], A. Fedorov[2], F. Ferri[7], D. Franci[15], G. Franzoni[36], K. Freudenreich[25], W. Funk[23], S. Ganjour[7], S. Gascon[9], M. Gataullin[38], F. X. Gentit[7], A. Ghezzi[14,23], A. Givernaud[7], S. Gninenko[19], A. Go[26], B. Gobbo[17], N. Godinovic[3], N. Golubev[19], P. Govoni[14], N. Grant[28], P. Gras[7], M. Haguenauer[8], G. Hamel de Monchenault[7], M. Hansen[23], J. Haupt[36], H. F. Heath[28], B. Heltsley[34], W. Hintz[25], R. Hirosky[33], P. R. Hobson[31], A. Honma[23], G. W. S. Hou[27], Y. Hsiung[27], M. Huhtinen[23], B. Ille[9], Q. Ingram[24], A. Inyakin[36], P. Jarry[7], C. Jessop[37], D. Jovanovic[22,a], K. Kaadze[35], V. Kachanov[21], S. Kailas[13], S. K. Kataria[13], B. W. Kennedy[29], P. Kokkas[11], T. Kolberg[37], M. Korjik[2], N. Krasnikov[19], D. Krpic[22,a], Y. Kubota[36], C. M. Kuo[26], P. Kyberd[31], A. Kyriakis[10], M. Lebeau[23], P. Lecomte[25], P. Lecoq[23], A. Ledovskoy[33], M. Lethuillier[9], S. W. Lin[27], W. Lin[26], V. Litvine[38], E. Locci[7], E. Longo[15], D. Loukas[10], P. D. Luckey[25], W. Lustermann[25], Y. Ma[38], M. Malberti[14], J. Malclès[7], D. Maletic[22], N. Manthos[11], Y. Maravin[35], C. Marchica[24,25], N. Marinelli[37], A. Markou[10], C. Markou[10], M. Marone[16], V. Matveev[19], C. Mavrommatis[10], P. Meridiani[23], P. Milenovic[22,25], P. Miné[8], O. Missevitch[2], A. K. Mohanty[13], F. Moortgat[25], P. Musella[18], Y. Musienko[19,32], A. Nardulli[25], J. Nash[30], P. Nedelec[6], P. Negri[14], H. B. Newman[38], A. Nikitenko[30], F. Nessi-Tedaldi[25], M. M. Obertino[16], G. Organtini[15], T. Orimoto[38], M. Paganoni[14], P. Paganini[8], A. Palma[15], L. Pant[13], A. Papadakis[5], I. Papadakis[10], I. Papadopoulos[11], R. Paramatti[15], P. Parracho[18], N. Pastrone[16], J. R. Patterson[34], F. Pauss[25], J-P. Peigneux[6], E. Petrakou[10], D. G. Phillips II[33], P. Piroué[39], F. Ptochos[5], I. Puljak[3], A. Pullia[14], T. Punz[25], J. Puzovic[22,a], S. Ragazzi[14], S. Rahatlou[15], J. Rander[7], P. A. Razis[5], N. Redaelli[14], D. Renker[24], S. Reucroft[32], P. Ribeiro[18], C. Rogan[38], M. Ronquest[33], A. Rosowsky[7], C. Rovelli[15], P. Rumerio[23], R. Rusack[36], S. V. Rusakov[20], M. J. Ryan[30], L. Sala[25], R. Salerno[14], M. Schneegans[6], C. Seez[30], P. Sharp[23,30], C. H. Shepherd-Themistocleous[29], J. G. Shiu[27], R. K. Shivpuri[12], P. Shukla[13], C. Siamitros[31], D. Sillou[6], J. Silva[18], P. Silva[18], A. Singovsky[36], Y. Sirois[8], A. Sirunyan[1], V. J. Smith[28], F. Stöckli[25], J. Swain[32], T. Tabarelli de Fatis[14], M. Takahashi[30], V. Tancini[14], O. Teller[23], K. Theofilatos[25], C. Thiebaux[8], V. Timciuc[38], C. Timlin[30], M. Titov[7], A. Topkar[13], F. A. Triantis[11], S. Troshin[21], N. Tyurin[21], K. Ueno[27], A. Uzunian[21], J. Varela[18,23], P. Verrecchia[7], J. Veverka[38], T. Virdee[23,30], M. Wang[27], D. Wardrope[30], M. Weber[25], J. Weng[25], J. H. Williams[29†], Y. Yang[38], I. Yaselli[31], R. Yohay[33], A. Zabi[8], S. Zelepoukine[21,25], J. Zhang[36], L. Y. Zhang[38], K. Zhu[38] and R. Y. Zhu[38]

[1] *Yerevan Physics Institute, Yerevan, Armenia*
[2] *Research Institute for Nuclear Problems, Minsk, Belarus*
[3] *Technical University of Split, Split, Croatia*
[4] *University of Split, Split, Croatia*
[5] *University of Cyprus, Nicosia, Cyprus*
[6] *Laboratoire d'Annecy-le-Vieux de Physique des Particules, IN2P3-CNRS, Annecy-le-Vieux, France*
[7] *DSM/DAPNIA, CEA/Saclay, Gif-sur-Yvette, France*
[8] *Laboratoire Leprince-Ringuet, Ecole Polytechnique, IN2P3-CNRS, Palaiseau, France*
[9] *Institut de Physique Nucléaire de Lyon, Université Lyon 1, CNRS/IN2P3, Villeurbanne, France*
[10] *Institute of Nuclear Physics ``Demokritos'', Aghia Paraskevi, Greece*
[11] *University of Ioánnina, Ioánnina, Greece*
[12] *University of Delhi, Delhi, India*
[13] *Bhabha Atomic Research Centre, Mumbai, India*
[14] *Istituto Nazionale di Fisica Nucleare e Università degli Studi di Milano Bicocca, Milano, Italy*
[15] *Sapienza Università di Roma e Sezione dell'INFN, Roma, Italy*
[16] *Università di Torino e Sezione dell'INFN, Torino, Italy*
[17] *Università di Trieste e Sezione dell'INFN, Trieste, Italy*
[18] *Laboratório de Instrumentação e Física Experimental de Partículas, Lisboa, Portugal*
[19] *Institute for Nuclear Research, Moscow, Russia*
[20] *Lebedev Physical Institute, Moscow, Russia*
[21] *IHEP, Protvino, Russia*
[22] *Vinca Institute of Nuclear Sciences, Belgrade, Serbia*
[23] *CERN, European Organisation for Nuclear Research, Geneva, Switzerland*
[24] *Paul Scherrer Institut, Villigen, Switzerland*
[25] *Institute for Particle Physics, ETH Zurich, Zurich, Switzerland*
[26] *National Central University, Chung-Li, Taiwan*
[27] *National Taiwan University (NTU), Taipei, Taiwan*
[28] *University of Bristol, Bristol, United Kingdom*
[29] *Rutherford Appleton Laboratory, Didcot, United Kingdom*
[30] *Imperial College, University of London, London, United Kingdom*
[31] *Brunel University, Uxbridge, United Kingdom*
[32] *Northeastern University, Boston, Massachusetts, USA*
[33] *University of Virginia, Charlottesville, Virginia, USA*
[34] *Cornell University, Ithaca, New York, USA*
[35] *Kansas State University, Manhattan, Kansas, USA*
[36] *University of Minnesota, Minneapolis, MN, USA*
[37] *University of Notre Dame, Notre Dame, IN, USA*
[38] *California Institute of Technology, Pasadena, California, USA*
[39] *Princeton University, Princeton, NJ, USA*
[40] *Florida State University, Tallahassee, FL, USA*
[a] *Also at Faculty of Physics of University of Belgrade, Belgrade, Serbia*
[b] *Now at the Institut für Synchrotronstrahlung - ANKA, Forschungszentrum Karlsruhe, Germany*
[†] *Deceased*

[*] *E-mail:* ioan.dafinei@roma1.infn.it

## Abstract

Ensuring the radiation hardness of $PbWO_4$ crystals was one of the main priorities during the construction of the electromagnetic calorimeter of the CMS experiment at CERN. The production on an industrial scale of radiation hard crystals and their certification over a period of several years represented a difficult challenge both for CMS and for the crystal suppliers. The present article reviews the related scientific and technological problems encountered.

# 1 Introduction

The Compact Muon Solenoid (CMS) [1] is a general purpose detector installed at the Large Hadron Collider (LHC) at CERN, Geneva. Detection and precise energy measurement of photons and electrons is a key to new physics that is expected at the 100 GeV - TeV scale. The discovery of the postulated Higgs boson is a primary goal at LHC and $H \rightarrow \gamma\gamma$ is the most promising discovery channel if the mass is between 114 and 130 GeV. In this mass range the Higgs decay width is very narrow, but the signal will lie above an irreducible background and so good energy resolution is crucial. A photon energy resolution of 0.5% above 100 GeV has therefore been set as a requirement for the CMS performance.

The CMS experiment has opted [2] for a hermetic homogeneous electromagnetic calorimeter (ECAL), made of lead tungstate ($PbWO_4$) crystals. The choice of lead tungstate has been driven by operating conditions which require that the ECAL be fast and highly granular and be able to withstand radiation doses of up to 4 kGy and $4\cdot10^{13}$ n/cm$^2$ in the Barrel and 25 times more in the Endcaps. These doses correspond to an integrated luminosity of 500 fb$^{-1}$ expected to be accumulated over 10 years.

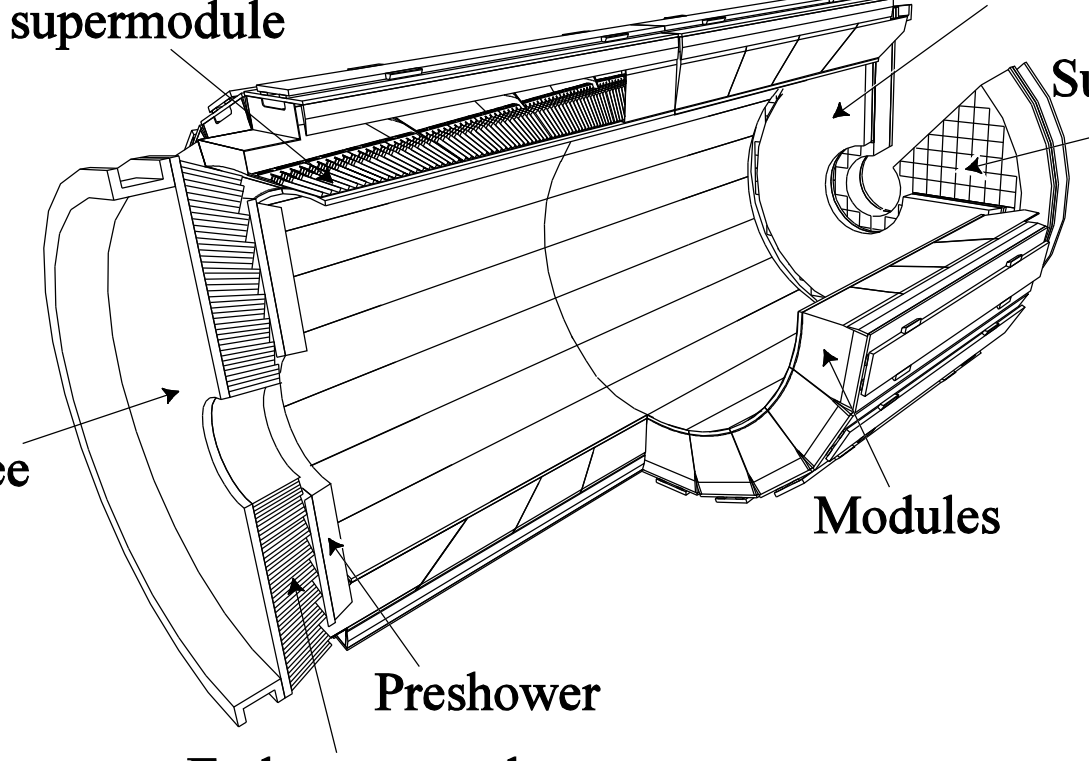

Fig. 1.1 ECAL layout.

The ECAL layout is shown in Fig.1.1. The high density of lead tungstate allows the design of a very compact detector. The region in space covered by a particle detector is usually described by pseudorapidity η, which is a spatial coordinate related to the angle of a particle relative to the beam axis. It is defined as:

$$\eta = -\ln\left[\tan\left(\frac{\Theta}{2}\right)\right] \tag{1.1}$$

where θ is the angle between the particle momentum and the beam axis. The ECAL consists of a Barrel, covering the pseudorapidity range $|\eta| < 1.479$, and two Endcaps which extend the coverage up to $|\eta| = 3$. There are 61200 crystals in the Barrel and 7324 crystals in each of two Endcaps, amounting to more than 91 metric tons of $PbWO_4$ crystals, with a total volume of approximately 11 m$^3$. The Barrel crystals are slightly tapered with dimensions about 24x24x230 mm$^3$. The exact shape varies with the pseudorapidity, requiring 17 different geometrical types each one having two symmetries (left and right). Following dedicated studies aimed at optimizing the uniformity of the light yield, one of the lateral faces of the Barrel crystals is semi-polished (roughness Ra ~ 0.25 μm) while all other faces are optically polished (roughness Ra < 0.02 μm). The Endcap crystals are less tapered, all identical in shape and have dimensions of about 30x30x220 mm$^3$. All faces of Endcap crystals are optically polished (roughness Ra < 0.02 μm).

The design energy resolution of CMS requires important properties of the $PbWO_4$ crystals, in particular:

- a large enough light yield (LY) to keep the stochastic contribution to the energy resolution small,
- a uniform longitudinal response to avoid a large constant term in the energy resolution at high energy, induced by the longitudinal fluctuations of electromagnetic showers.

These properties must be maintained in the high radiation field mentioned above. Furthermore, one should be able to track and correct for any radiation-induced changes in the light yield at the level of a few tenths of a percent. The aim of this paper is to present how these issues have been studied and solved by the CMS ECAL group in a way compatible with an unprecedentedly large industrial production. The paper is organized as follows: Section 2 describes the radiation environment of ECAL. Section 3 presents the general properties of the lead tungstate scintillator with particular emphasis on its radiation hardness related properties. The experimental methods and tools used to characterize $PbWO_4$ crystals for CMS and related experimental results are presented in sections 4 and 5 respectively. Section 6 is dedicated to the monitoring of the light yield variation under irradiation conditions. Finally, section 7 describes the practical implementation of the quality assurance for the mass production for the two producers, Bogoroditsk Technochemical Plant (BTCP) in Russia and Shanghai Institute of Ceramics, Chinese Academy of Sciences (SICCAS) in China.

# 2 ECAL radiation environment

Operating at a peak luminosity of $10^{34}$ $cm^{-2}s^{-1}$, the LHC will produce a very harsh radiation environment for the detectors. The ECAL will be exposed to fast hadrons, mostly pions, which, in interactions with the ECAL itself, produce secondary hadrons, and build up a flux of low energy neutrons, with energies typically below 10 MeV. In addition, electromagnetic showers inside the crystals produce a significant dose.

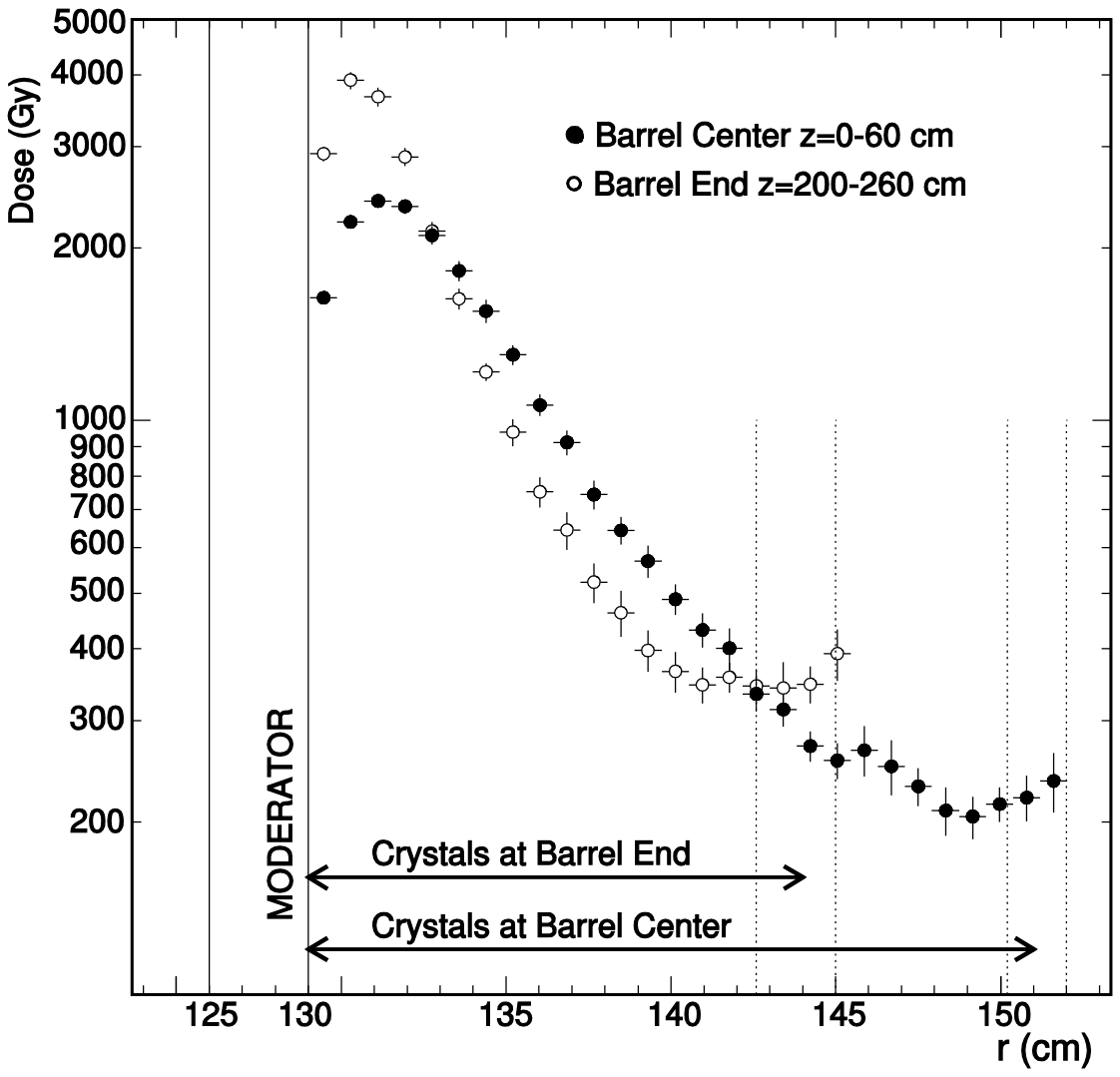

Fig. 2.1 Absorbed dose as a function of radial distance from the beam axis in the center and at the ends of the Barrel ECAL [2]. Averaging has been performed over 60 cm in z, which causes the end of the crystal envelope to span the range indicated in the plot by the vertical dotted lines. Values are for 500 $fb^{-1}$ integrated luminosity.

No single quantity is sufficient to fully characterize this complex environment. A fairly complete picture can be obtained by looking at the absorbed gamma-irradiation dose, the neutron fluence and the charged hadron fluence. With respect to radiation damage, threshold behavior is often observed, i.e. only particles above a certain energy cause damage. These aspects will be discussed in detail in the sections 5.2 and 5.3 of this article, but it appears that low-energy (< 20MeV) neutrons do not cause significant damage in the crystals. As for damage due to fast hadrons, it has been suggested [3] that it is better parametrized in terms of the density of their inelastic interactions than by the hadron fluence directly.

Therefore we introduce three quantities to characterize the environment:

- The absorbed dose
- The density of inelastic hadronic interactions (star density)
- The neutron fluence below 20 MeV

For consistency, the values are presented for an integrated luminosity of 500 $fb^{-1}$, expected to be reached after 10 years, or $5 \cdot 10^7$ s operation at peak luminosity. Thus the average dose per hour at $10^{34}$ $cm^{-2}s^{-1}$ will be $7.2 \cdot 10^{-5}$ of the 10-year integral. However, while the LHC is being filled and the beams accelerated there is no associated radiation, while during the subsequent collision period of typically 20 h duration, the luminosity will decrease by a factor of about 5 from its initial value. Thus there will be substantial short-term variations in the instantaneous dose rate.

Fig. 2.1 shows the absorbed dose in the Barrel ECAL as a function of radial distance from the beam axis. It can be seen that the value varies by roughly an order of magnitude over the crystal length. Note that the data are plotted as a function of radius, not along the (tilted) crystal axis.

Fig. 2.2 shows, as a function of pseudorapidity, the three radiation quantities we use to characterize the environment: absorbed dose, star density and neutron fluence. The solid symbols in Fig. 2.2 correspond to their average values along the crystal axis and the open symbols correspond to the maximum values. In the Barrel all three quantities are almost independent of $\eta$. Here the dose rate at shower maximum corresponds to 0.17–0.25 Gy/h at nominal LHC peak luminosity. However, in the Endcaps, the radiation levels increase rapidly with increasing $\eta$. The performance requirements are most stringent for the region below $|\eta| = 2.5$, which is the limit of coverage of the central tracking detector. This limit is indicated by the dotted vertical line in figure 2.2, where it can be seen that, at this value of $\eta$, the dose rate at shower maximum is 5 Gy/hr. An overall uncertainty of about a factor of 2 should be assigned to the dose estimates, which are obtained from Monte Carlo simulations [2].

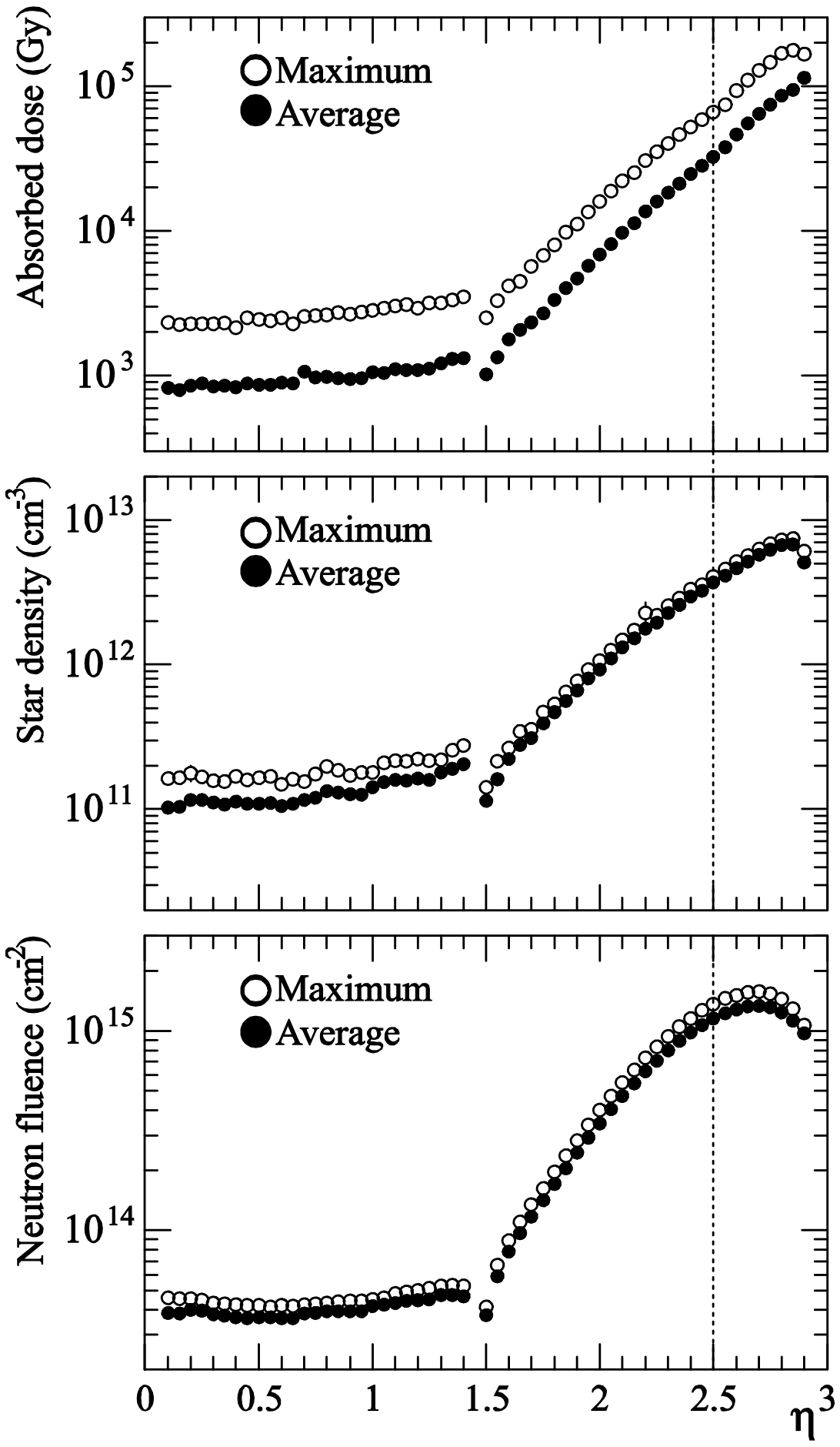

Fig. 2.2 Radiation exposure of the ECAL as a function of $\eta$. The star densities are for interactions above 20 MeV hadron energy and neutron fluences comprise neutrons with energies between thermal and 20 MeV. The dotted vertical line indicates the limit of coverage of the CMS central tracker (see text). All values are for 500 fb$^{-1}$ integrated luminosity.

# 3 General properties of $PbWO_4$ scintillator

In order to meet the granularity and energy resolution requirements, CMS selected lead tungstate ($PbWO_4$) as the most suitable affordable scintillator for its electromagnetic calorimeter. This crystal combines a number of attractive properties, such as high density, fast luminescence, good radiation tolerance when properly optimized [4-10] and adequate light yield. Table 3.1 summarizes the main physical and optical properties of $PbWO_4$. The scintillation emission may be described as the sum of three exponential terms with the time constants (relative intensities) shown in column six of the table. Lead tungstate is bi-refringent; column eight gives values for the two refractive indices (ordinary, $n_o$ and extraordinary, $n_e$) at two wavelengths (420 nm and 600 nm).

Table 3.1: Lead tungstate crystal properties

| $\rho$, g/cm$^3$ | $X_0$, cm | Molière radius, cm | LY, ph/MeV | Temperature dependence of LY [11] %/°K | $\tau_{sc}$, ns (%) | $\lambda_{em}$, nm | Refractive indices [12] $n_o/n_e$ |
|---|---|---|---|---|---|---|---|
| 8.28 | 0.89 | 2.19 | 200 | -1.98 | 5 (73%)<br>14 (23%)<br>110 (4%) | 420 | 420nm : 2.360/2.240<br>600nm : 2.242/2.169 |

Considerable efforts over many years were made to understand the properties and to optimize the performance of $PbWO_4$ for the demanding CMS specifications. The methods needed to grow large and colorless crystals, to suppress slow scintillation components and to reach good radiation tolerance levels were investigated. In addition the development of mass production technologies (crystal growth and mechanical processing) allowing this unprecedented production to be made with a high production yield was addressed with high priority. A review of this work and of $PbWO_4$ properties is given in [10]. A detailed description of the $PbWO_4$ growth method applied at SICCAS is given in [13].

## 3.1 Requirements of radiation resistance

All known crystal scintillators suffer from radiation damage. The most common damage is radiation-induced light absorption caused by the formation of color centers. The absorption coefficient is proportional to the density of color centers which in most cases is proportional to the concentration of defects and impurities pre-existent in the crystal. Radiation may also cause phosphorescence (afterglow), which leads to increased readout noise. Additional effects may include reduced scintillation light yield (damage to the scintillation mechanism) and a change in the light response uniformity (since the radiation dose profile is usually not uniform). Light output variation caused by radiation-induced absorption can be corrected by external light monitoring [1]. On the other hand, a variation of the scintillation mechanism or a strong non-uniformity of the light response cannot be monitored in situ. This sets therefore two major requirements for the CMS application: the scintillation mechanism must not be affected by radiation and the density of radiation-induced color centers must be kept below a maximum level.

## 3.2 Nature of point defects and luminescence centers

The crystallographic structure of synthetic lead tungstate crystal has been determined by X-ray diffraction and identified as sheelite-type with tetragonal symmetry and space group $I4_1/a$. The parameters of the unit cell are $a = b = 5.456$, $c = 12.020$ Ångström. However a significant loss of lead during the crystal growth process is expected to induce a reorganization of the lattice with the majority of lead and oxygen related vacancies distributed in a sheelite-like superstructure with a slightly reduced symmetry (space group $P\bar{4}$) and lattice constant values: a = b = 7.719, c = 12.018 Ångström [14]. In this structure the ordering of the vacancies is compensated by a distortion of the tungstate anionic polyhedra. However the existence of this superstructure has not been confirmed by neutron diffraction studies [15] nor by other dedicated X-ray diffraction studies [16-18]. Although the presence of an optimum of radiation resistance for crystals grown slightly off-stoechiometry seems to play in its favor, the cation vacancy ordering in a superstructure in $PbWO_4$ still remains a matter of discussion.

$PbWO_4$ crystals grown from purified raw materials (5N or 6N with Molybdenum contamination limited to a few ppm) have their scintillation and radiation damage properties completely dominated by cation and anion vacancies and not by impurities. The position of these vacancies, either as part of the scheelite-like structure or randomly distributed, may influence these properties.

It has been established in several studies (see [19] for detailed analysis and extended references) that the luminescence of $PbWO_4$ crystals is related to charge transfer transitions in the regular anionic molecular complex $WO_4^{2-}$ and in similar tetrahedra but distorted by an oxygen vacancy $WO_3$. The charge state of these centers not being modified by gamma and charged particle irradiation (at least in the range of doses and dose rates expected at LHC) the scintillation properties of $PbWO_4$ (radio- and photo-luminescence spectra, decay time, intrinsic light yield) should remain unchanged in the LHC radiation environment.

## 3.3 Radiation damage mechanism and point-defect compensation by specific doping

When $PbWO_4$ crystals are exposed to ionizing radiation pre-existing point-structure defects may act as traps for electrons or holes. The resulting charged defects have discrete energy levels and optical transitions can be induced, absorbing part of the scintillation light during its transport to the photodetector. This is the well known mechanism of radiation-induced color centers, which is the main source of optical damage in lead tungstate at LHC. Ionizing radiation damage can be considered as a three step process consisting of:

- creation of hot electrons and holes from the interaction of high energy particles with the lattice
- free carrier separation during the thermalization (through strong coupling with lattice phonons) and diffusion process
- localisation of electrons and holes near lattice defects, to balance local charge

Up to five types of color centers have been identified in $PbWO_4$ with corresponding absorption bands at 350-400, 470, 520, 620 and 715 nm. Through detailed studies of EPR (electron paramagnetic resonance), TSL (thermo-stimulated luminescence), PSC (photo-stimulated conductivity) and TSC (thermo-stimulated conductivity) it was possible to identify the corresponding color centers as follows:

- 350-400 nm: $WO_3^{2-}$ di-electron center on a Frenkel defect resulting from an out-of-position oxygen atom
- 470 and 520 nm: several types of di-$O^-$ centers
- 620 nm: electron transfer from the valence band to the lead vacancy ($V_c$) related defect $O^-V_cO^-$ ground state
- 715nm: photo-ionization of the dimer center $(WO_3 + WO_3)^{2-}$

However, the induced absorption spectrum and relative intensity of these five bands strongly depend on the nature and density of pre-existing structural defects, which depend themselves on the crystal growth conditions. Undoped crystal grown from stoechiometric raw-material have an absorption spectrum with two dominating broad bands peaked near 380 and 620 nm.

The doping of $PbWO_4$ crystals by specified impurities such as La, Y, Nb at different stages of the growth process has been used for the production of CMS crystals to improve their radiation hardness. Indeed, pentavalent doping with niobium prevents the trapping of holes on oxygen near a lead vacancy by forcing oxygen leakage. Cation vacancies can be compensated by substituting stable trivalent ions for lead ions in the nearest co-ordination sphere around the defect. Different ions have been tried like $Y^{3+}$, $La^{3+}$, $Lu^{3+}$, $Gd^{3+}$, $Al^{3+}$. A very significant suppression of the electron/hole trapping processes is already observed for a doping concentration of the order of 100 ppm if the crystal stoechiometry is well controlled [5, 6, 10].

### 3.4 Kinetics of damage production and recovery

A parameter which characterises the change in optical properties induced by radiation exposure is the radiation-induced absorption coefficient (μ), defined as:

$$\mu = \frac{1}{L} \cdot \ln \frac{T_{init}}{T_{irr}} \tag{3.1}$$

where $T_{init}$ and $T_{irr}$ are respectively the values of the crystal transmission measured before and after irradiation and L is the length of the crystal along the measurement direction. For a given wavelength, the induced absorption coefficient is directly proportional to the total concentration of all color centers absorbing at this wavelength:

$$\mu \propto N = \sum_i N_i \tag{3.2}$$

At a given time, $t$, under continuous irradiation at a fixed dose rate S the radiation-induced absorption coefficient μ in the crystal results from the balance between the creation of color centers with damage constant $d_i$ (related to the cross section of free carrier capture by defects of type i) and their destruction due to annealing at the detector operating temperature with a recovery rate $\omega_i$. The process may be described by the following equation [20]:

$$\mu = \sum_i \frac{\mu_i^{sat} S}{S + \omega_i d_i} \left\{ 1 - \exp\left[ -\left( \omega_i + \frac{S}{d_i} \right) t \right] \right\} \tag{3.3}$$

where $\mu^{sat}_i$ is the induced absorption coefficient at saturation due to defects of type i. After some time a dynamic equilibrium is reached, which is dose rate dependent as expressed by the first term of the equation (3.3). In general, it is less than the saturation level,

$$\mu^{sat} = \sum_i \mu_i^{sat} = \sum_i N_i \sigma_i \tag{3.4}$$

where $N_i$ is the concentration of color centers of type i and $\sigma_i$ is their photon absorption cross section.

In ECAL working conditions, each collision run will be followed by a refill period. Supposing a constant irradiation rate S for a time interval $t_0$ (during collision run), the recovery of the induced absorption during refill time will be described by the equation:

$$\mu = \sum_i \frac{\mu_i^{sat} S}{S + \omega_i d_i} \left\{ 1 - \exp\left[ -\left( \omega_i + \frac{S}{d_i} \right) t_0 \right] \right\} \exp\left( -\omega_i (t - t_0) \right) \tag{3.5}$$

Under normal LHC operation conditions the dose rate will be such that the full defect saturation will never be reached, and the optical transmission of the crystals will oscillate around a point corresponding to a loss as compared to its initial value, which is less than a few percent except at large values of $\eta$ in the Endcaps. In order for these oscillations to remain as small as possible and to be able to monitor them at a reasonable frequency during the periods of beam and machine refill, crystals have been optimized so that fast recovery constants at room temperature (in the range of minutes) are kept as small as possible. In some cases this has led to an increase of slow recovery centers (SICCAS crystals), in other cases to an increase of ultrafast recovering (microseconds) shallow traps (BTCP crystals), which are not harmful for the CMS detector operation. This optimization has been made for the centers absorbing in the domain of the scintillation emission spectrum, i.e. for the 350-400, 470, 520 nm centers and is not necessarily valid for red absorbing centers, which generally have a faster recovery constant. The 420 nm induced absorption recovery is well fitted with a double exponential with time constants of about 1 hour and 40 to 75 hours respectively. The infrared band however has a much faster recovery of about 8 minutes; this characteristic allows monitoring at long wavelengths to be used as a cross-check of the performance of the laser monitoring system.

# 4 Experimental methods, measurement equipment and parameters

The investigation of radiation damage mechanisms in $PbWO_4$ involved a large number of experts in different scientific domains and complex testing equipment in many locations. In order to build a complete picture of radiation effects in $PbWO_4$ it was necessary to study all possible sources of damage, namely electromagnetic (gammas and electrons), charged hadrons and neutrons. It was therefore mandatory to organize irradiation facilities able to cover a large range of doses and dose rates for a detailed study of the crystal behavior in all possible conditions, even those with low probability, such as very high dose rates arising from accidental beam loss. Special care was taken to cross-check the dosimetry of irradiation facilities used for the certification of the mass production. The uniformity of dose through the volume of a crystal was of particular relevance. The irradiation facilities used by ECAL group are summarized in the Appendix.

The crystal radiation damage behavior in ECAL CMS was assessed through a number of optical measurements, as described below.

## 4.1 Optical transmission and LY measurements

To compute the induced absorption parameter, $\mu$, one has to measure the optical transmission at different times during or after irradiation exposure. Standard optical transmission measurement of $PbWO_4$ crystals used in CMS comprised measurements made along the crystal length (longitudinal transmission, LT) and transversely (transverse transmission, TT) in the direction having both lateral faces optically polished. A number of purpose built or commercial spectrophotometers were used in different laboratories. In most of the cases these were double beam spectrophotometers with 2 nm wavelength resolution or better. Scans were usually performed with a step of 5 to 10 nm and the transmission measurement precision was typically about 1%. During the ECAL production, transmission and LY measurements were performed with automatic systems ACCOS [21] at CERN and ACCOR [22] at the Italian National Institute for Nuclear Physics (INFN). A second, simplified version of an ACCOR machine (only transmission measurements) was also used for the irradiation tests made at INFN. In addition, a dedicated spectrophotometer [23] was built at INFN for high precision measurements of the optical properties of $PbWO_4$ crystals (reflection, absorption, refractive index, surface quality and geometrical effects). The instrument was also used for high precision measurements of radiation-induced absorption coefficient in $PbWO_4$ crystals.

The LY is defined as the number of photoelectrons produced in a photomultiplier placed at the rear face of the crystal for 1 MeV of energy deposited (pe/MeV). A standard CMS measurement for crystals consisted of measuring the LY for 22 positions of the radioactive source along the crystal, starting 0.5 cm from the front face, in 1 cm steps. Measured LY values are the result of a fit procedures described in [21] and [24] for the two automatic machines ACCOS and ACCOR respectively. The LY value given in this work corresponds to the value calculated at 8 $X_0$ (7.5 cm from the front face) using the formula:

$$LY_{8X_0} = a + 7.5 \cdot b \qquad (4.1)$$

where a and b are respectively the intercept and slope of the linear fit of 11 LY values starting from the front face of the crystal.

Studies of a number of $PbWO_4$ crystals have shown that LY and longitudinal transmission are strongly correlated [25]. As will be discussed in the next section, the scintillation mechanism is not affected by radiation. This characteristic implies a strong correlation between LY loss and the induced absorption coefficient at wavelengths close to the scintillation emission peak (400-450 nm), an essential condition for a reliable monitoring of the crystal yield with injected light pulses.

### 4.2 Scintillation spectrum analysis

Scintillation spectrum analysis allows investigation of the nature of light emission center(s) and the possible influence of radiation exposure [26-29]. In particular the photo-luminescence and radio-luminescence spectra were recorded over the whole temperature range from liquid helium temperature to 200°C, before and after irradiation with dedicated (purpose built) or commercial spectrofluorimeters such as the Perkin-ELMER LS 55 at CERN. In the case of radio-luminescence the crystal was excited by a pulsed X-ray source (Hitachi) or in a synchrotron radiation facility (Hasylab in Hamburg or LURE in Orsay).

### 4.3 Decay time measurement

Another technique used in the study of radiation-induced traps was investigation of the decay time of the different emission centers as a function of temperature before and at different times after irradiation. Usually these tests were performed in parallel with those mentioned in paragraph 4.1 above. The setup was based on the single delayed photon counting method originated by Bollinger and Thomas [30].

### 4.4 Other analytical methods

In addition to the optical characterization of crystal defects, which was done systematically in the early stage of the R&D, and also on a sampling basis during all the production period, a number of other analytical methods were used such as X-ray diffraction, GDMS (Glow Discharge Mass Spectroscopy), ENDOR (Electron-Nuclear Double Resonance) spectroscopy, SEM (Scanning Electron Microscopy), ESR (Electron Spin Resonance), EPR (Electron Paramagnetic Resonance), TSL (Thermo-Stimulated Luminescence), TSC (Thermo-stimulated Conductivity) and PSC (Photo-stimulated Conductivity).

## 5 Radiation damage effects in $PbWO_4$ crystals

### 5.1. Radiation damage effects under gamma irradiation

A key point in the studies of $PbWO_4$ radiation damage mechanisms is to verify that the scintillation mechanism is not modified by the radiation (see section 3.1). This was shown in early work on $PbWO_4$ [11, 26, 27] and further studied by detailed analysis of the light emission centers. Figure 5.1.1 shows a comparison of the radio-luminescence (X-ray excited) spectra measured before and after 1 kGy (350Gy/h) gamma irradiation for a $PbWO_4$ crystal of the standard ECAL production. The shape of the luminescence spectrum is not changed by the gamma irradiation, indicating no damage to the scintillation mechanism.

In order to study the nature and relative concentration of gamma radiation-induced color centers, three types of transmission spectra were found to be useful as a means of classification [27]. In addition research was carried out aimed at establishing the microscopic origin of radiation damage in $PbWO_4$ (nature and properties of related color centers) and finding ways to improve the crystals radiation hardness [31-41]. Reliable certification procedures for crystals produced on an industrial scale for ECAL were defined on the basis of this R&D activity.

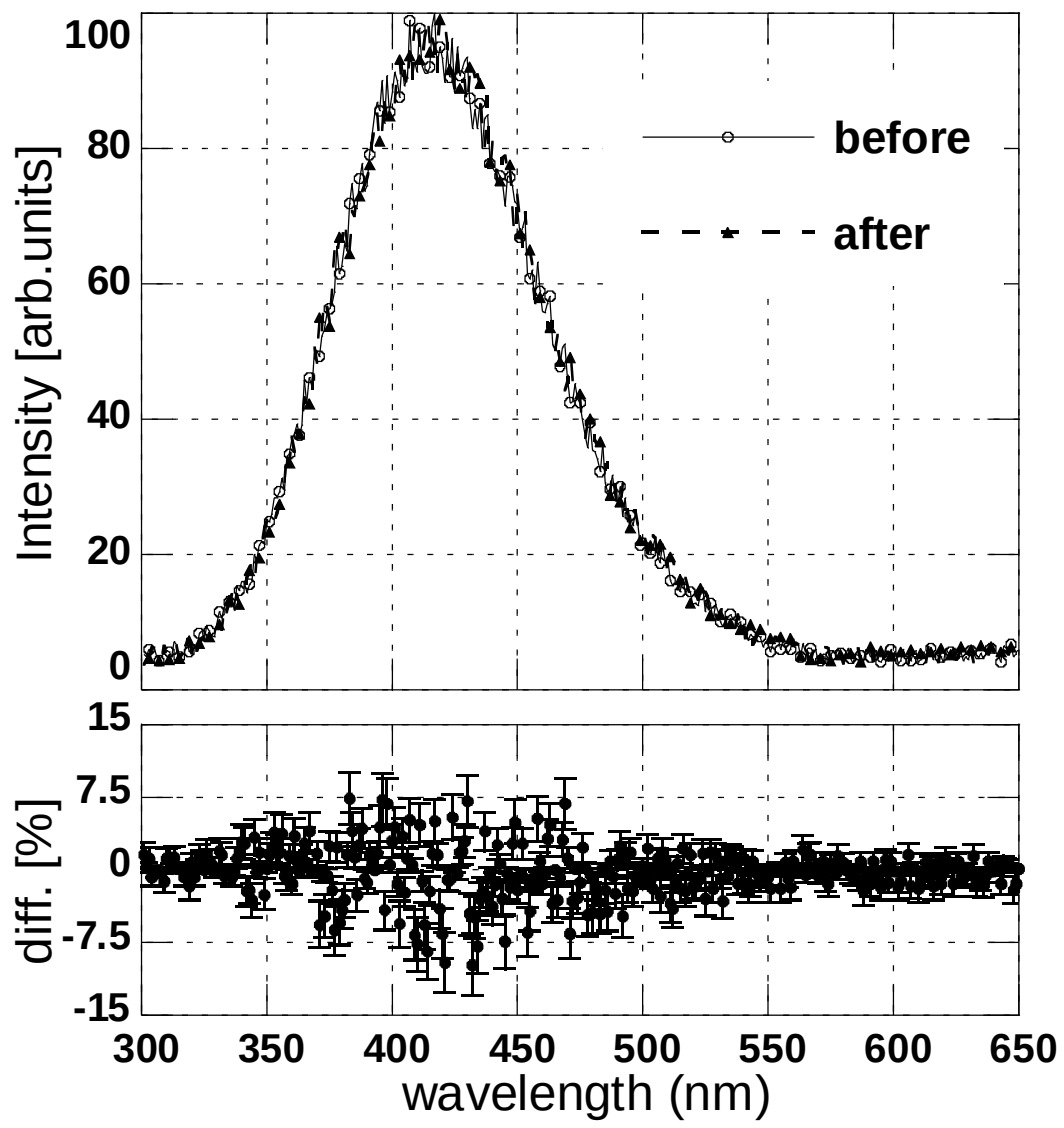

Fig. 5.1.1 Above: radio-luminescence (X-ray excited) spectrum of $PbWO_4$ crystal before and after gamma radiation exposure (1 kGy at 350 Gy/h). Below: the difference between the two measurements.

A large number of radiation hardness tests were performed on different production crystals. Unless otherwise stated, the results reported here employed lateral irradiation (simultaneous irradiation of the full length of the crystal from the side) with a $^{60}$Co source. These tests showed that radiation-induced color centers are always the same. Only their relative concentration and their distribution along the crystal growth axis may differ depending on the raw material characteristics (stoechiometry and possible doping), crystal growth conditions and post-growth thermal treatments. All these production parameters were different for the two crystal suppliers, BTCP and SICCAS. Nevertheless dedicated production protocols and certification procedures were defined at both producers in order to have similar radiation hardness characteristics of the crystals. Fig. 5.1.2 shows typical absorption coefficient ($\mu_{LT}$) induced by gamma radiation, measured along the length of the crystal, in randomly selected production crystals.

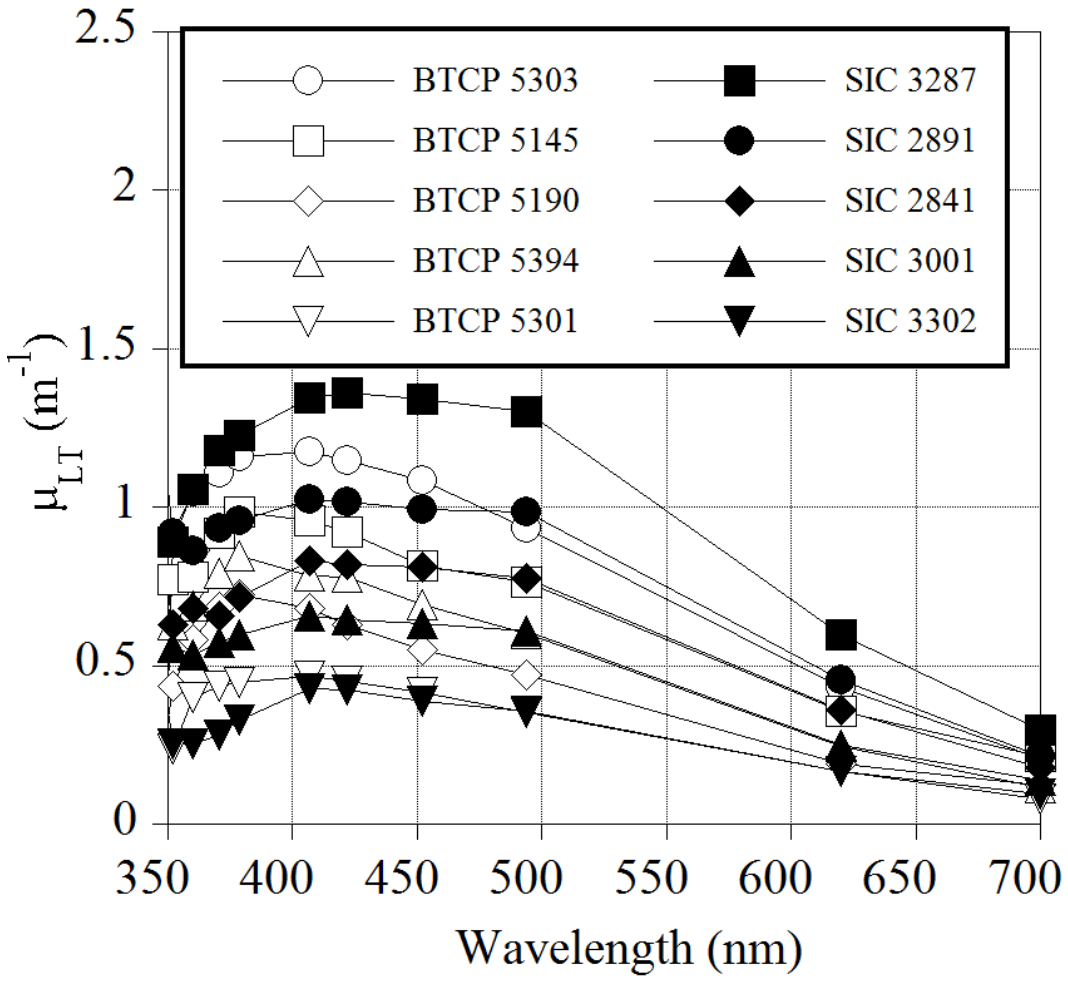

Fig. 5.1.2 Radiation-induced absorption coefficient as a function of wavelength measured over the length of a crystal on randomly selected production crystals. Results were obtained after a total dose of 350 Gy, at a dose rate of 350 Gy/h. Open and solid symbols refer to BTCP and SICCAS crystals respectively.

Fig. 5.1.3 shows the value of the induced absorption coefficient $\mu_{TT}$ at 420 nm measured transversely at different points along the growth axis of the crystals. The different variation of color centers along the crystal growth axis due to different growth technique (Czochralski at BTCP, Bridgman at SICCAS) may be noted. The increase in color centers in Bridgman-grown crystals along the axis is compensated by appropriate cutting of the ingot such that the front region of the crystal (most exposed to irradiation in LHC conditions) has the lowest concentration of color centers (highest radiation hardness).

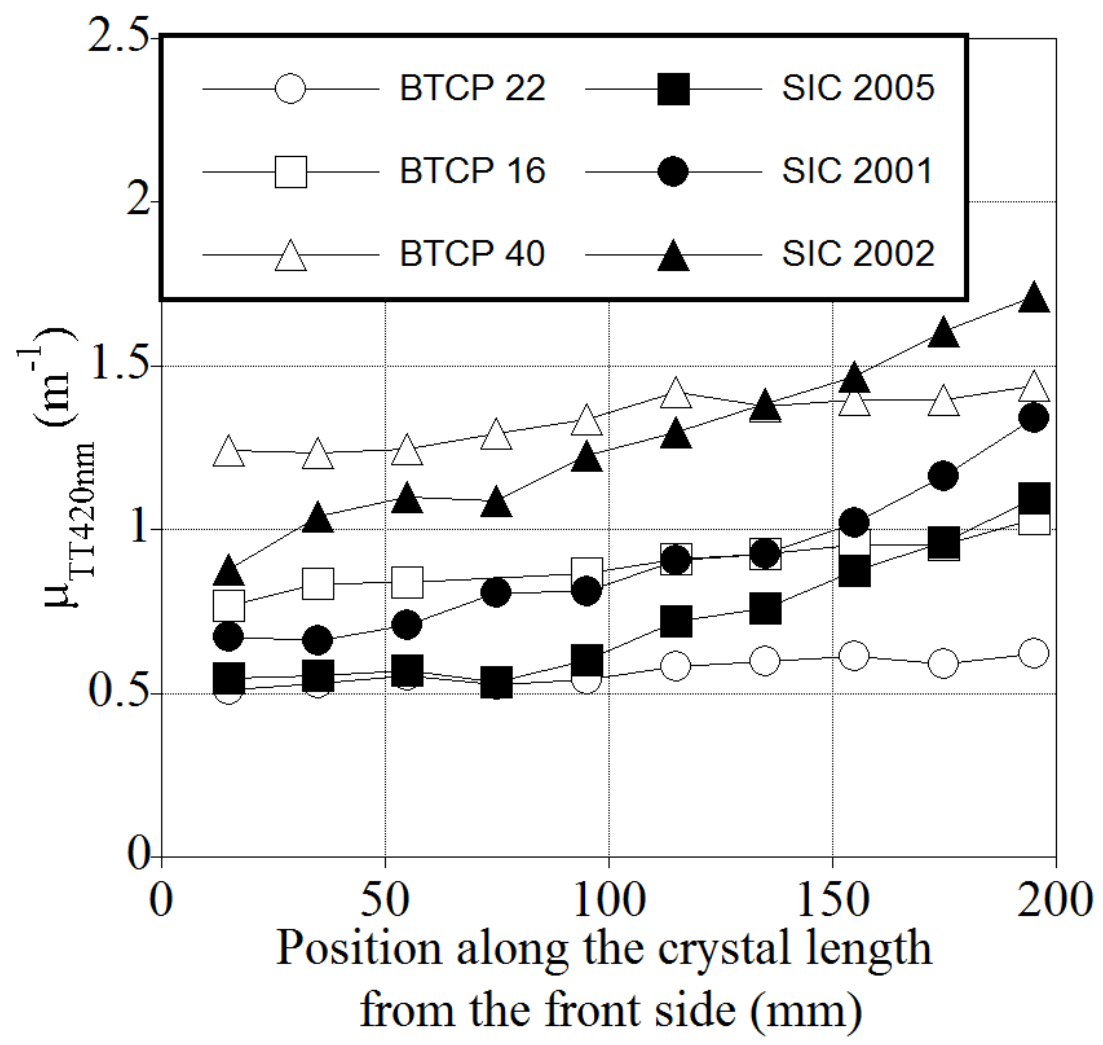

Fig. 5.1.3 Radiation induced absorption coefficient at 420 nm measured transversely at different points along the growth axis of randomly selected production crystals. Results were obtained after a total dose of 350 Gy, at a dose rate of 350 Gy/h. Open and solid symbols refer to BTCP and SICCAS crystals respectively.

Studying the distribution of the damage along the growth axis may give valuable information to the crystal grower such as the distribution of defects responsible for color center production with respect to the seed position, stability of growth conditions and quality of the raw material.

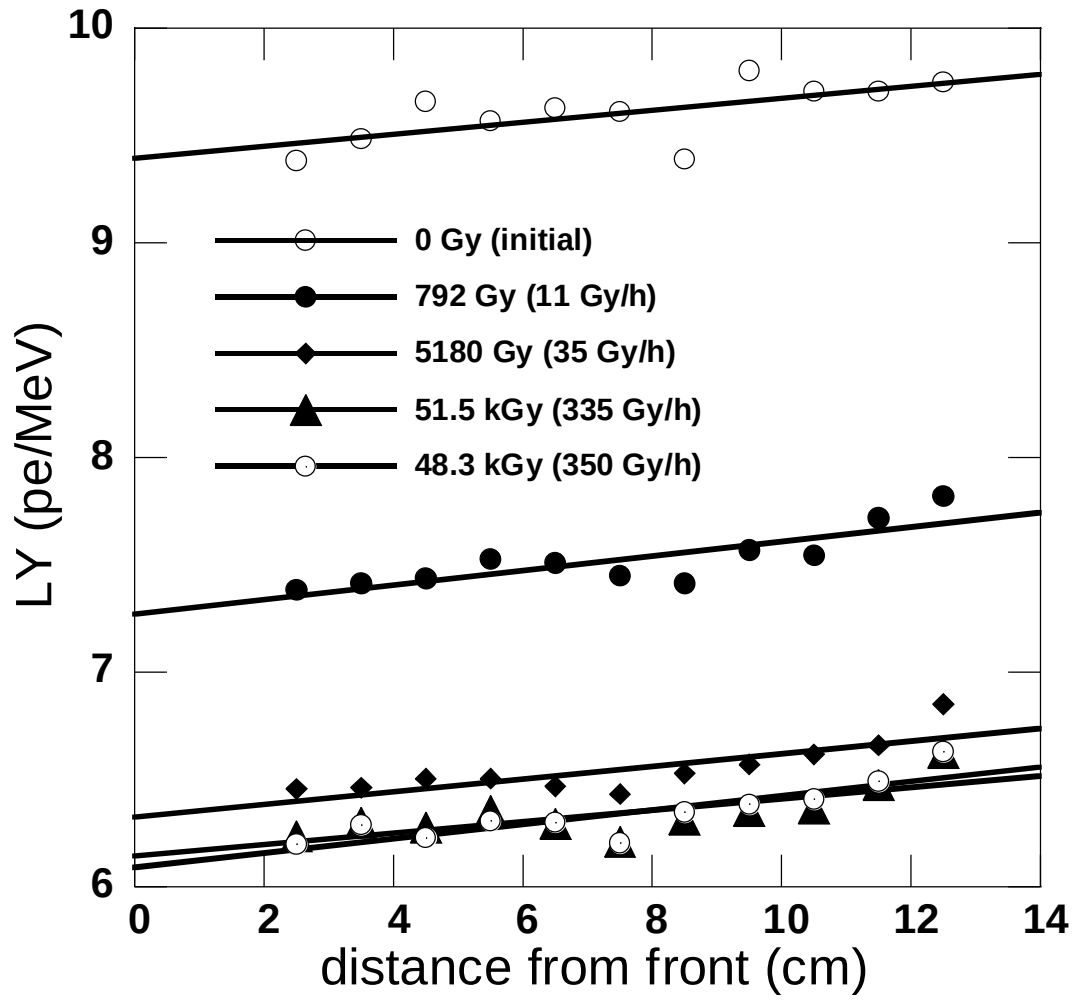

Fig. 5.1.4 Light yield uniformity measured under equilibrium conditions for different gamma radiation doses for a randomly selected production crystal. The lines are linear fits to the data.

Fig. 5.1.4 shows the LY uniformity under gamma irradiation of a production crystal. The measurement was made using a collimated source positioned at different points along crystal. Measurement was made after reaching the equilibrium of induced absorption, with dose rates of 11 Gy/h, 35 Gy/h, 335 Gy/h and 350 Gy/h. Linear fits show that, within measurement errors, the slope of the LY uniformity, 0.03 pe/MeV/cm is not changed under gamma ray dose rates up to 350 Gy/h and integrated doses of several tens of kGy.

Based upon such studies and on Monte Carlo simulations of the effect of optical absorption on energy resolution, the maximum acceptable value of the induced absorption coefficient $\mu_{LT}$ at saturation was set at 1.5 $m^{-1}$ for BTCP crystals and 1.6 $m^{-1}$ for SICCAS crystals at the peak emission wavelength (420nm), which corresponds to a light yield loss of approximately 6% (see also section 6 and refs. [42, 43]).

Radiation-induced color centers anneal at room temperature, leading to spontaneous recovery of radiation damage. Fig. 5.1.5 shows an example of spontaneous recovery at room temperature.

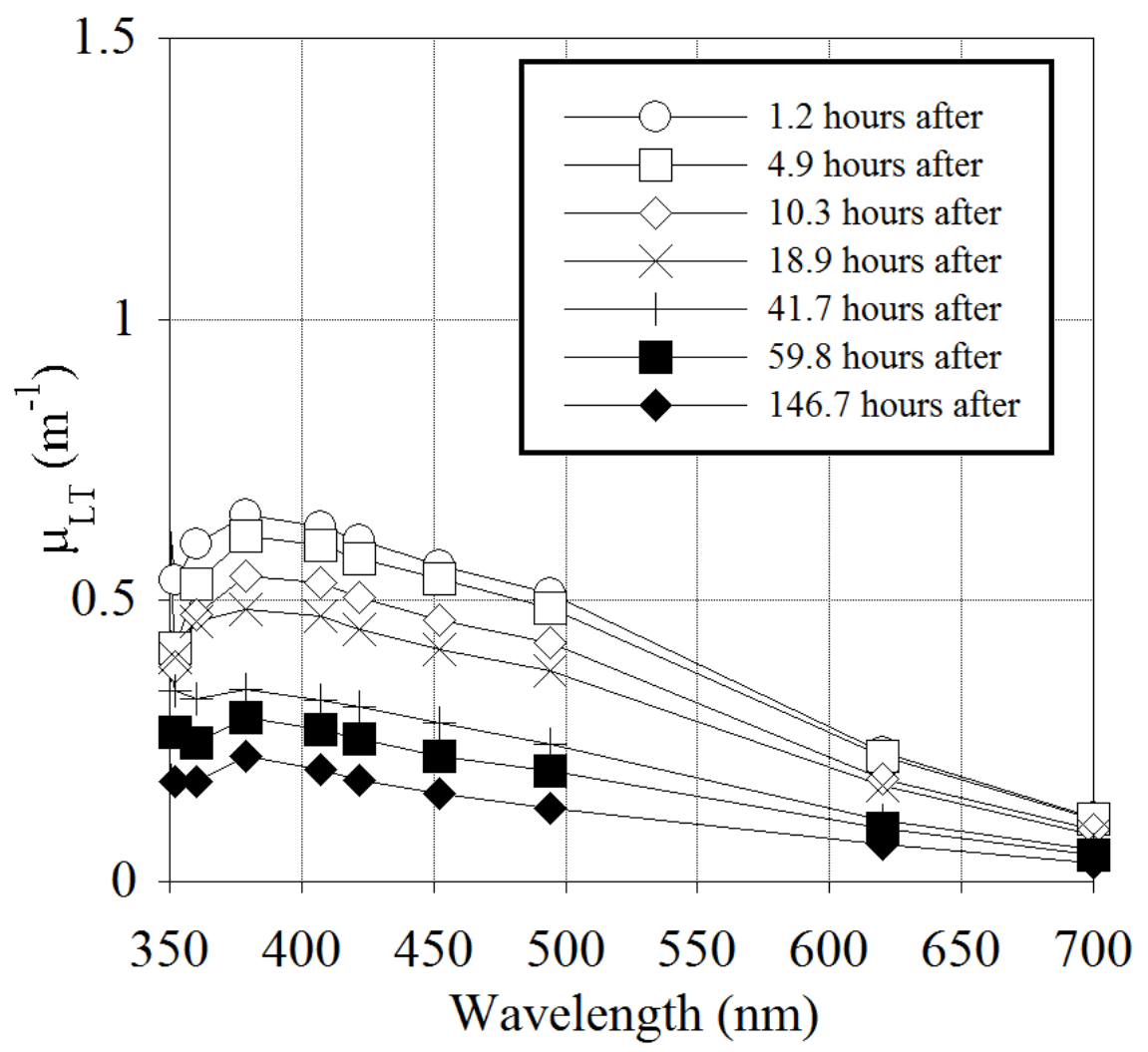

Fig. 5.1.5 Induced absorption coefficient in a crystal measured at different time after gamma irradiation at a rate of 350 Gy/h for 1 h. Recovery in the dark at room temperature (20°C).

In common with other scintillating crystals, thermal annealing and optical bleaching were found effective in removing radiation-induced absorption in $PbWO_4$ crystals [28]. Thermal annealing at 200°C for two hours was found to be effective in removing all residual color centers, and was used as a standard procedure for restoring crystals to their initial condition following gamma irradiation.

### 5.1.1 Dose rate dependence and low dose rate irradiation effects

The kinetics of color center creation under ionizing radiation and spontaneous annealing at room temperature may be described by simple models [19, 39], valid in the case of a low concentration of crystal defects. Such simple models lead to a two exponential time dependence of the concentration of color centers (i.e. of the induced absorption coefficient). Different studies on the interplay between gamma radiation damage and its recovery confirmed that the radiation-induced color center density depends on the dose rate. For a given dose a constant level is reached corresponding to the equilibrium between damage and recovery processes. Beyond a given dose rate and accumulated dose the induced absorption saturates corresponding to the full saturation of defects. Fig. 5.1.6 shows degradation of the longitudinal transmission of a production crystal for different gamma dose rates. Longitudinal transmission spectra were measured when the radiation damage reached equilibrium for a given dose rate.

Using longitudinal transmission data, e.g. Fig. 5.1.6, numerical values of radiation-induced absorption coefficients (related to the radiation-induced color center density) can be calculated. The shape of the induced absorption as a function of wavelength is the same for all crystals, indicating an identical nature of the radiation damage as shown by several studies [4-10, 29, 44].

The radiation dose in CMS will not add significantly to the number of point defects acting as color centers [45]. The main source of deep traps in $PbWO_4$ crystals are pre-existing oxygen related Frenkel defects and cation vacancies. However, synthetic lead tungstate crystals also contain shallow electron traps associated with oxygen vacancies. The long duration of the crystal irradiation creates conditions for the diffusion of the neutral shallow electronic traps significantly increasing the probability of their coupling in pairs and more complex defects. Pairs have deeper capture levels compared to single vacancies and can create metastable color centers increasing the optical absorption in the visible spectral region under ionizing radiation. The presence of such metastable color centers may explain the slow decrease of the crystal transmission (and consequently of measured light yield) and the existence of a dynamic saturation (equilibrium between damage and recovery) whose level depends on the radiation dose rate.

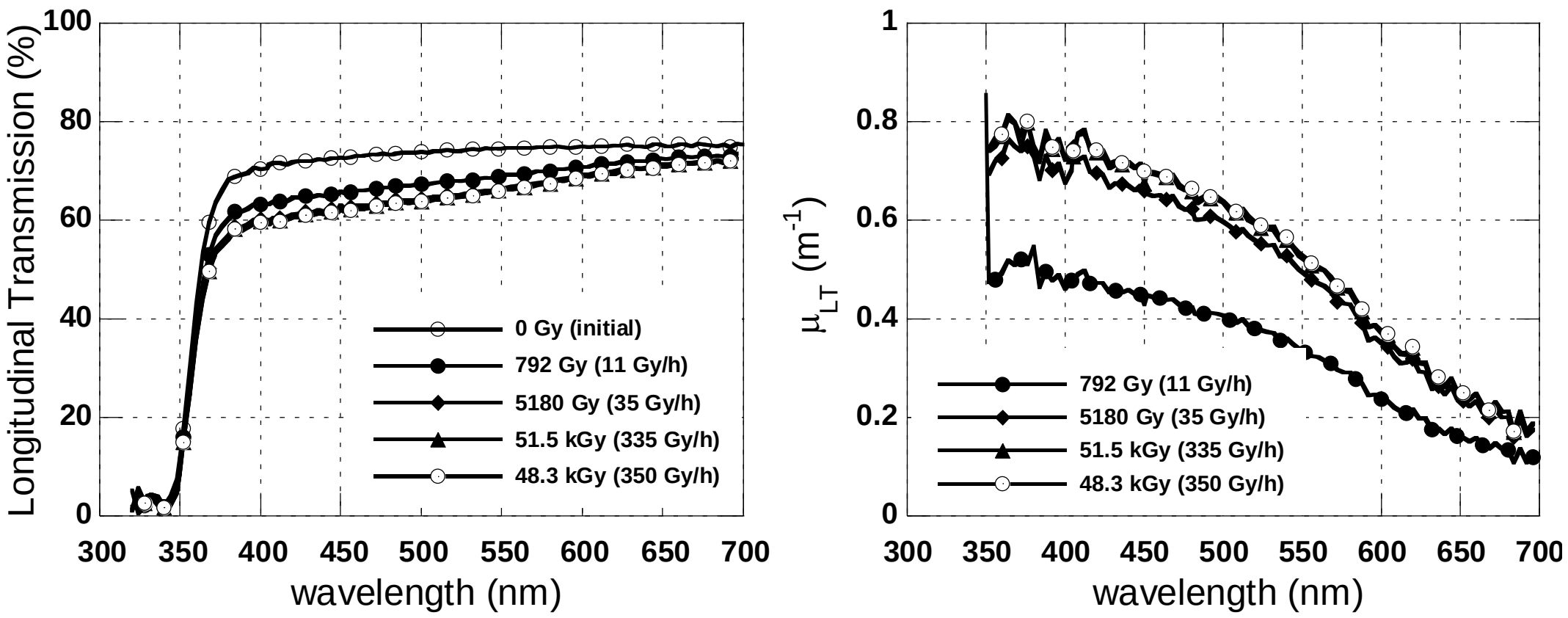

Fig. 5.1.6 Longitudinal transmission and corresponding equilibrium values for the induced absorption coefficient for a production crystal exposed to gamma irradiation. The measured transmission is not corrected for reflection at crystal faces. The crystals were thermally annealed before each irradiation cycle.

The characteristics of point defects responsible for color centre creation are also discussed in [37]. The parameters of trap levels responsible for the thermal bleaching of color centers are measured and the kinetics of the bleaching process is explained in the frame of a model which takes into account the annihilation of electron and hole centers by tunneling. The model clarifies the existence of the very slow recovery component which cannot be explained by simple thermal annealing.

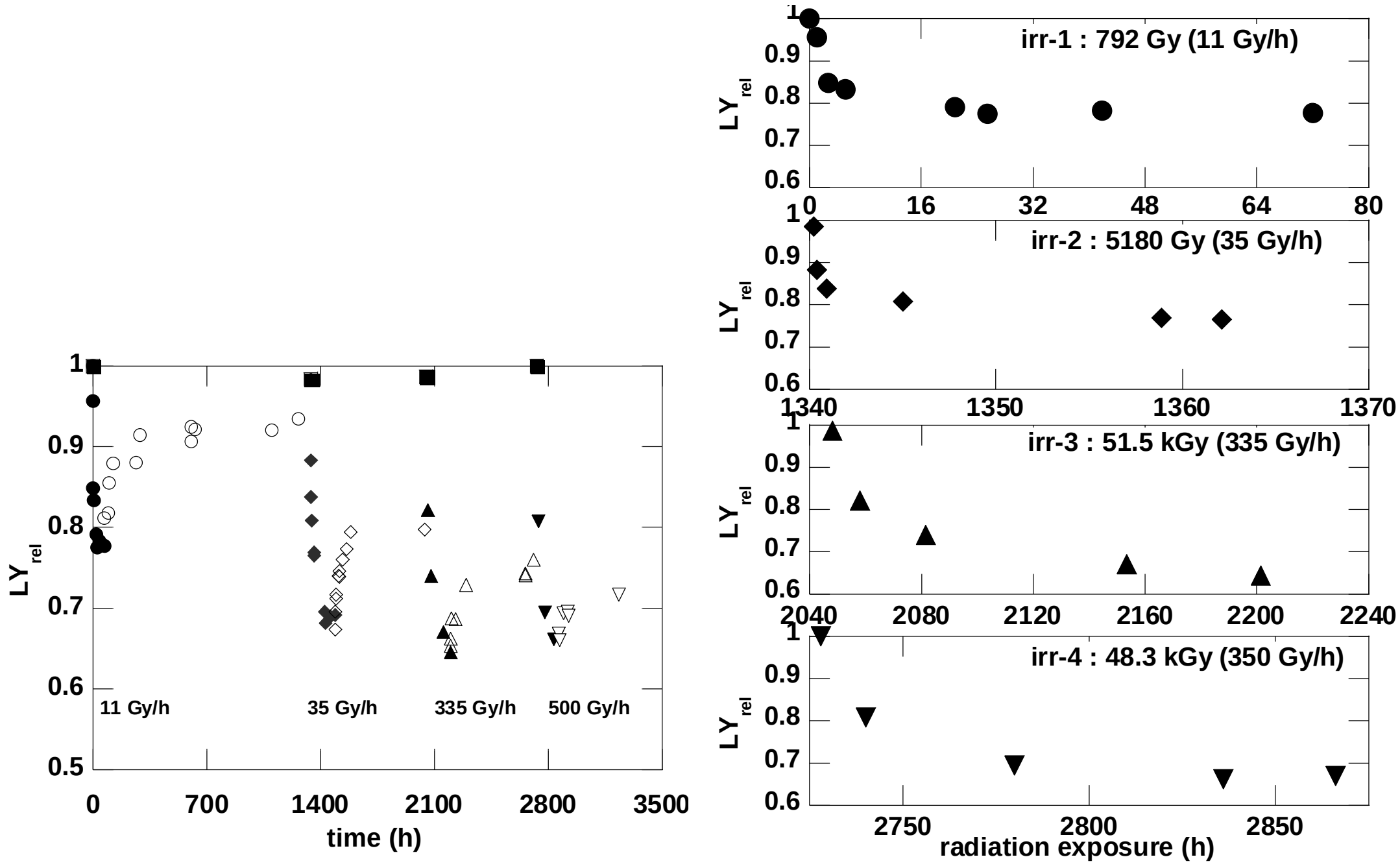

Fig. 5.1.7 Variation of light yield during gamma irradiation (solid symbols) and successive room temperature recovery (open symbols) of a production crystal (note different time scales of zoomed plots on the right side). Solid black squares are the values measured after the thermal annealing made before each irradiation cycle.

Fig. 5.1.7 shows the measured relative LY (normalized to the initial value, $LY_{rel}=LY/LY_{init}$) as a function of time under lateral gamma irradiation for a production crystal subjected to increasing dose rates of 11 Gy/h, 35 Gy/h, 335 Gy/h and 350 Gy/h. The radiation exposure continued at a given dose rate till the equilibrium condition of the induced absorption coefficient was reached, after which the recovery of damage at room temperature was observed for several hours. The saturation effect is better seen on the right side of the figure where different time scales are used for each irradiation step. Crystals were thermally annealed before each irradiation cycle. As seen in the figure, the LY degradation shows a clear dose rate dependence and, as for the induced absorption coefficient, different equilibrium values are reached for each radiation dose. Similar tests made on production crystals exposed at the dose rate expected for the CMS Barrel calorimeter in situ at LHC (0.15 Gy/h) showed a loss of the LY of less than 6% (see also section 7). Series production crystals exposed to LHC-like radiation conditions (front irradiation) presented a light yield loss below the CMS specification limit of 6% [42, 43]. The light yield of crystals from both suppliers shows similar dose rate dependent damage.

Fig. 5.1.8 shows crystal light yield variation measured under front irradiation described in [46] at a dose rate of 0.15 Gy/h. Crystals were irradiated for more than one month ($5 \cdot 10^4$ min) and a correction of the data for the irradiation source decay has been applied. The continuing decrease in light yield at long irradiation times can be attributed to the formation of a second type of color center which reaches equilibrium more slowly, as mentioned above

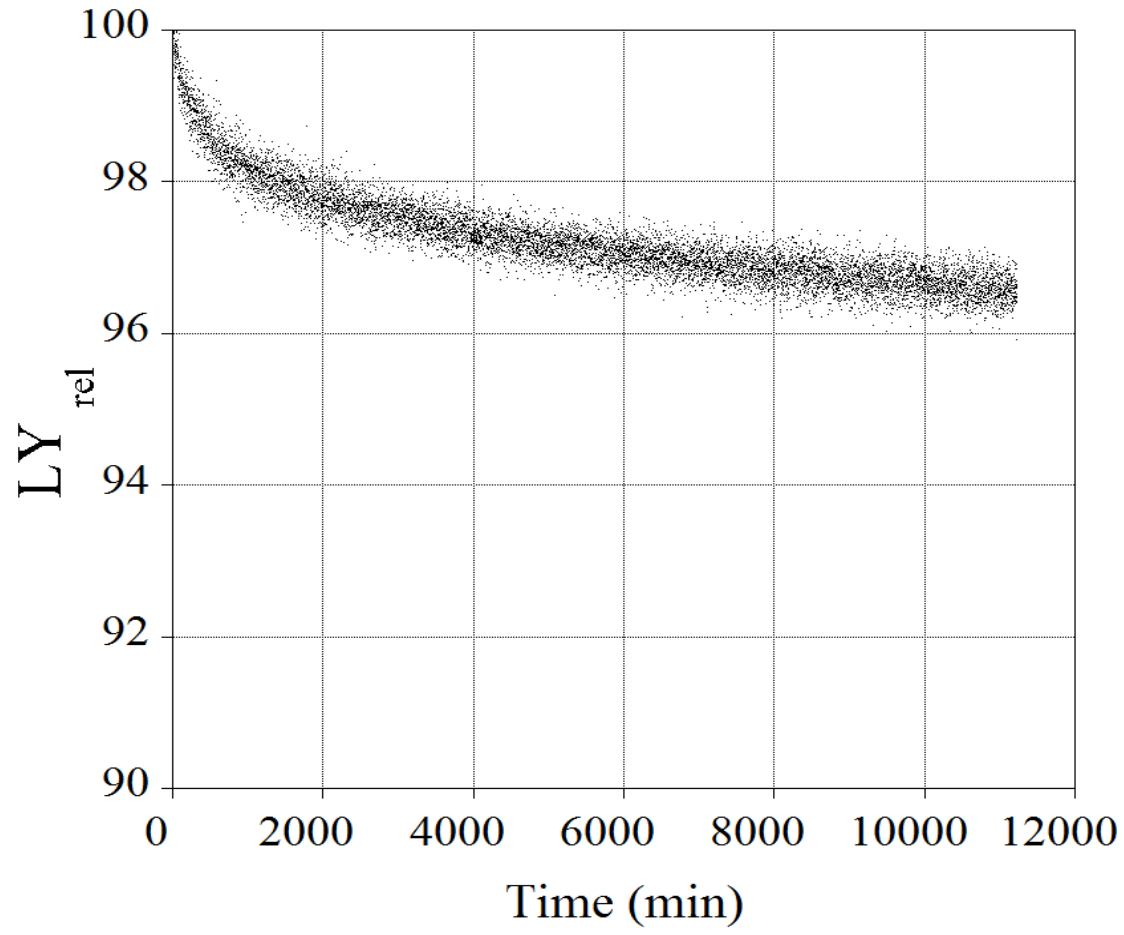

Fig. 5.1.8 Relative LY ($LY_{rel}$) is shown as a function of time for a production crystal under front irradiation with a dose rate of 0.15 Gy/h. The spread of measured values indicates the reproducibility of the measurements, which is better than 1%.

## 5.2. Radiation damage effects under neutron irradiation

In view of the intense neutron flux expected in CMS (see section 2) the effects on lead tungstate of neutron exposure were studied in nuclear reactors [47, 48]. The neutron fluxes and energies in these exposures were comparable to those expected in CMS. However, in reactors there is a strong associated gamma dose. The effect arising from neutrons was estimated by comparing the reactor results with results obtained from pure gamma irradiations. This indicated that there was no specific effect due to neutrons on the optical and scintillating properties of lead tungstate, at least up to fluences of $10^{14}$ $cm^{-2}$. This was confirmed by later independent studies [49]. It is also to be mentioned that recent tests performed at a very high fluence, of the order of $10^{19}$ to $10^{20}$ $n \cdot cm^{-2}$ and 330 MGy (i.e. well above the level that will be ever achieved in any physics experiment) revealed the robustness of lead tungstate crystals which were not destroyed nor locally vitrified, and remained scintillating after such heavy irradiation [50].

## 5.3. Radiation damage effects due to fast hadrons

In CMS the flux of fast hadrons is dominated by charged pions with energies of order 1 GeV. The effect of charged hadrons on crystals has been the subject of an extensive series of measurements, whose results are presented in Refs. [3, 51, 52]. Those studies show that hadrons cause a specific, cumulative loss of light transmission in the crystals [3]. The data for absorption induced by charged hadrons versus wavelength exhibit a $\lambda^{-4}$ dependence, unlike those for damage from γ-irradiations. As explained in [3], this is an indication of Rayleigh scattering from small centers of severe damage, as might be caused by the high energy deposition, along their path, of heavily ionizing fragments, locally generating extended defects in the crystal. Since the crystal contains heavy elements, fast-hadron specific damage is in fact expected from the production, above a ~20 MeV threshold, of heavy fragments with up to 10 μm range and energies up to ~100 MeV, causing a displacement of lattice atoms and energy losses along their path, up to 50000 times that of minimum-ionizing particles.

Apart from the effect on light transmission, no fast-hadron specific damage to the scintillation mechanism was observed within the accuracy of the measurements and for the explored flux and fluence ranges. This is evident from the correlations between LY loss and induced absorption, which show similar behavior for proton and gamma irradiated crystals [51].

The studies published in [3] and [51] were all performed with 20 GeV/c protons, thus leaving open the question how they should be rescaled to the much softer spectrum of pions expected in CMS at the LHC. This issue was studied in [52], where the damage from protons and from 290 MeV/c pions is compared. There it is shown that the pion-to-proton ratio of induced absorption coefficients is in agreement with the corresponding ratio of the density of inelastic interactions (stars) obtained from simulations. Thus, the average star densities expected in CMS at various values of pseudorapidity $\eta$, as given in Table 2.1 of Section 2 and in [3] can be used, together with the observed damage as a function of proton fluence in [3], to extract an expectation for the damage as shown in Fig. 5.3.

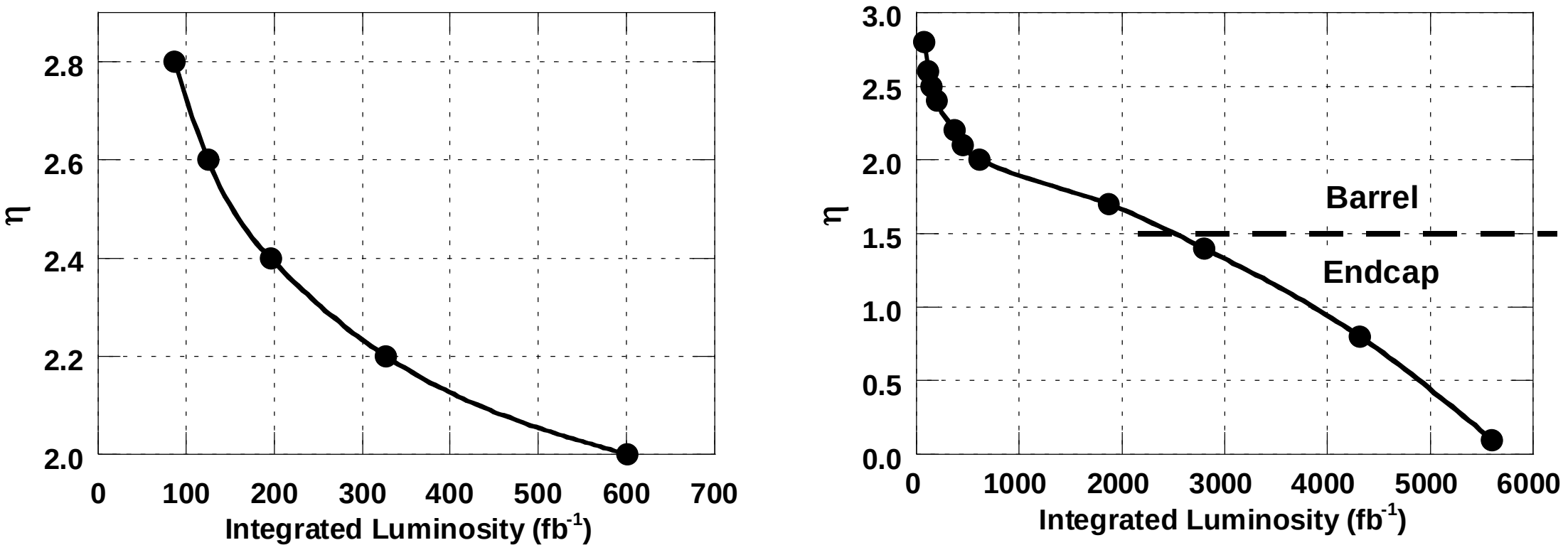

Fig. 5.3.1 Values for $\eta$ where an induced absorption coefficient 2 $m^{-1}$ due to charged hadrons is anticipated at the scintillation emission peak wavelength, as a function of integrated luminosity. The lines through the data points are to guide the eye.

However it should be pointed out that the induced absorption values used to obtain Fig. 5.3 were measured 150 days after irradiating the crystals, allowing the recovery of a damage component observed to have a time constant of 17.2 days [3]. With the data available in [3], the remaining damage could be fitted as a combination of stable damage (which is cumulative with fluence) and a recovering component with a time constant of 650 days. Thus the expectation for charged-hadron induced damage of Fig. 5.3.1 could change slightly if the recovery of the slow component were taken into account.

# 6 Monitoring of LY degradation and recovery under irradiation

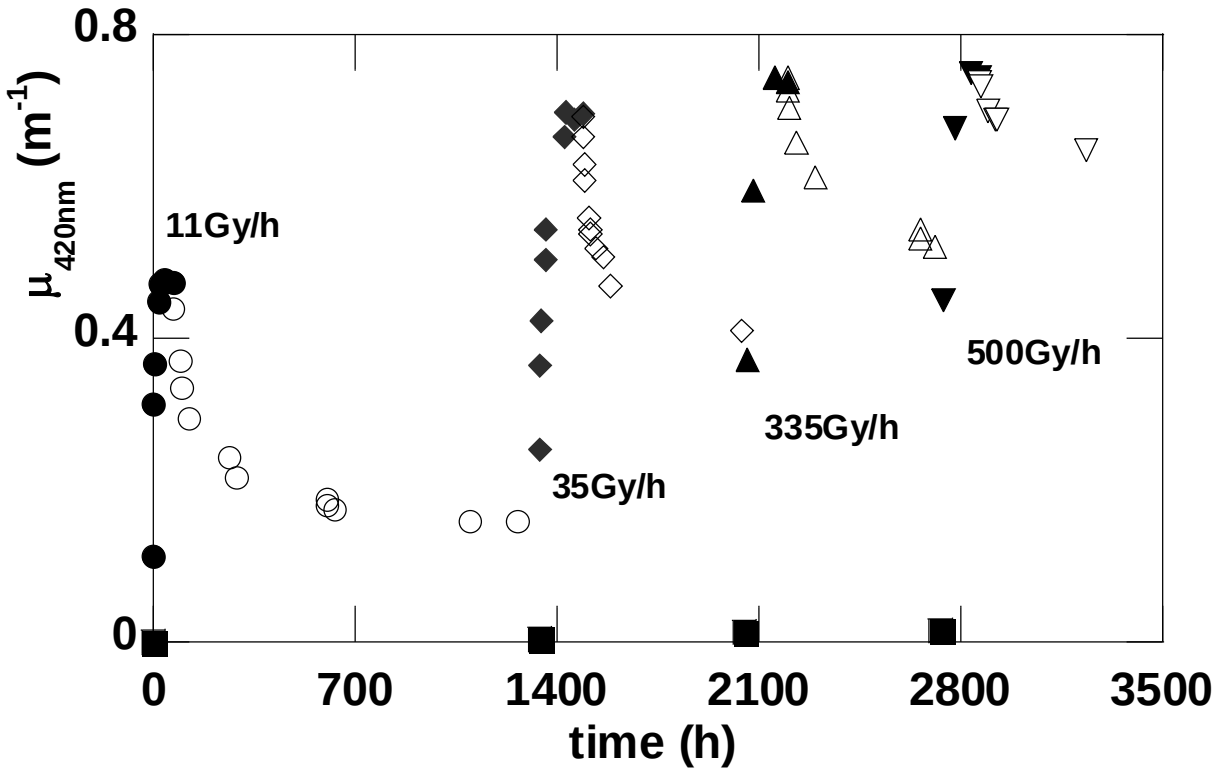

Fig. 6.1 Time evolution of the induced absorption coefficient at 420 nm during gamma irradiation and successive room temperature recovery of a production crystal. Solid and open symbols correspond to irradiation and recovery respectively. Solid black squares are the values measured after thermal annealing.

To avoid a degradation in the energy resolution, changes in the LY due to radiation induced absorption must be monitored during LHC operation. Variation of LY during irradiation exposure and spontaneous recovery cycles illustrated in fig. 5.1.7 are caused only by color center formation and annihilation and can therefore be monitored by tracking the corresponding modification of the optical transmission of the crystal. The induced absorption coefficient follows the same evolution as can be seen in Fig. 6.1 where the time evolution of the induced absorption coefficient at 420 nm measured longitudinally for the same crystal for which the LY time evolution is reported in fig. 5.1.7.

Fig. 6.2 shows the correlation between the light yield and the longitudinal transmission at 420 nm for the data presented in figures 5.1.7 and 6.1. The strong correlation demonstrates that variations in light yield can be corrected for by monitoring the changes in longitudinal transmission in the region of the emission peak.

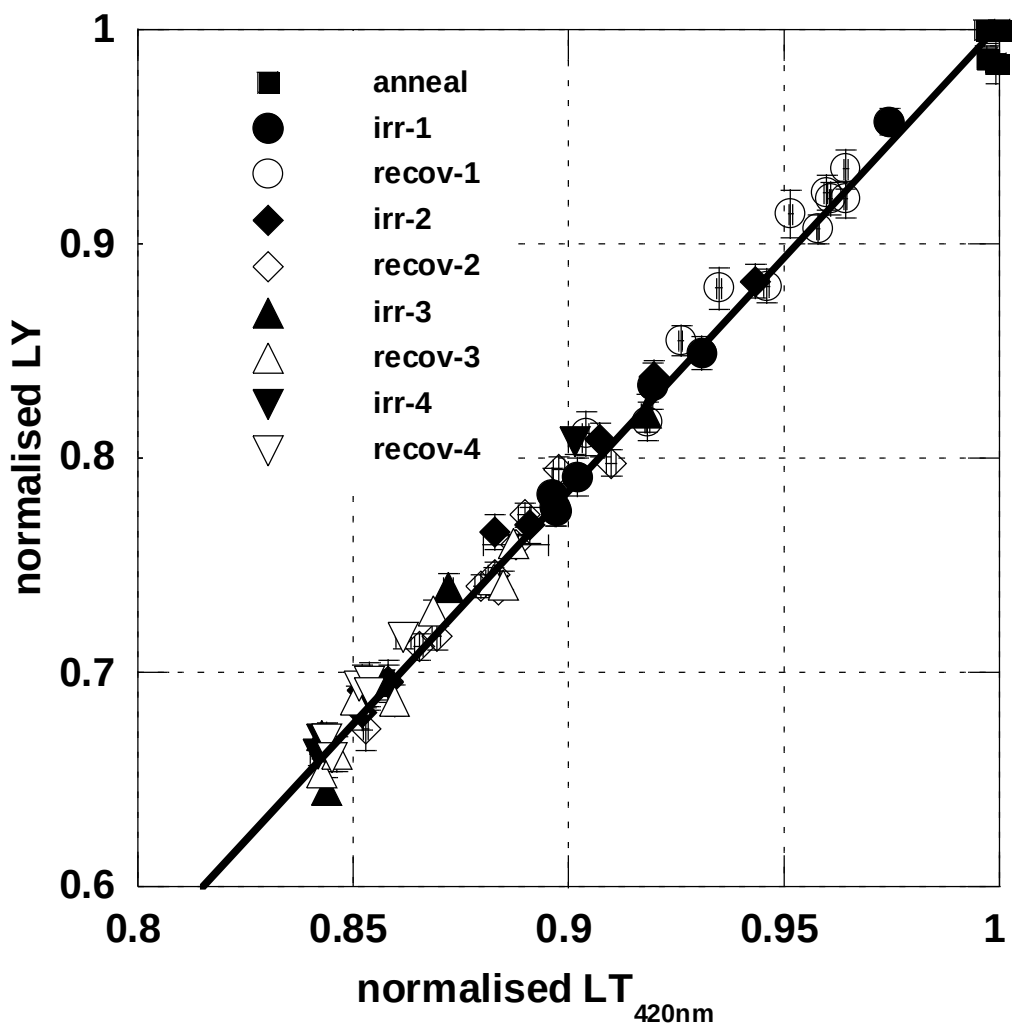

Fig. 6.2 Correlation between normalised LY and normalized longitudinal transmission during several radiation exposure-recovery cycles, for a production crystal. The line is a linear fit.

To take advantage of this correlation, a light monitoring system was designed and implemented [53-57]. This system was extensively used during high intensity tests in electron beams, showing that the energy resolution can be maintained by applying laser monitoring corrections under LHC-like conditions.

# 7 Crystal quality assurance during production

## 7.1. Crystal quality specification

It has been stressed throughout this paper that ensuring the radiation hardness of the crystals was of primary importance for CMS. Besides specifications related to dimension and optical properties needed for the required calorimeter performance, additional specifications for radiation hardness were defined for crystals produced at both crystal suppliers. These specifications fixed the acceptable limit of optical transmission damage and related LY deterioration as follows.

1) induced absorption for full saturation of crystal: $\mu < 1.5\ m^{-1}$ at 420 nm; irradiation conditions: lateral $^{60}Co$ source exposure for a total dose > 500 Gy (> 100 Gy/h)
2) light yield loss: $LY_{loss} < 6\%$; irradiation conditions: front $^{60}Co$ source exposure for a total dose > 2 Gy (> 0.15 Gy/h)

The first specification was set up in order to prevent the total damage exceeding a certain level when the radiation damage in the crystal is fully saturated throughout its volume. This level corresponds to a defect density estimated to be $3{\cdot}10^{17}$ color centers per $cm^3$, corresponding to an induced absorption of $1.5\ m^{-1}$ at the peak emission wavelength (420 nm) at full saturation. The limits imposed for irradiation conditions are the results of statistical studies made on production crystals showing that in general, the damage saturation is reached after a lateral gamma radiation exposure with a dose of at least 500 Gy at a rate above 10 Gy/h. The maximum value of $1.5\ m^{-1}$ placed on the induced absorption at the emission wavelength also limits an associated variation in the longitudinal uniformity of light collection which could degrade the energy resolution of the calorimeter.

The second specification is related to LY loss due to gamma radiation exposure close to the LHC irradiation conditions (0.15 Gy/h at high luminosity at the region of shower maximum in the crystal, about 3 cm from the crystal front face). Statistical studies made on production crystals show that LY degradation is saturated after a front $^{60}$Co source exposure for a total dose > 2 Gy (> 0.15 Gy/h).

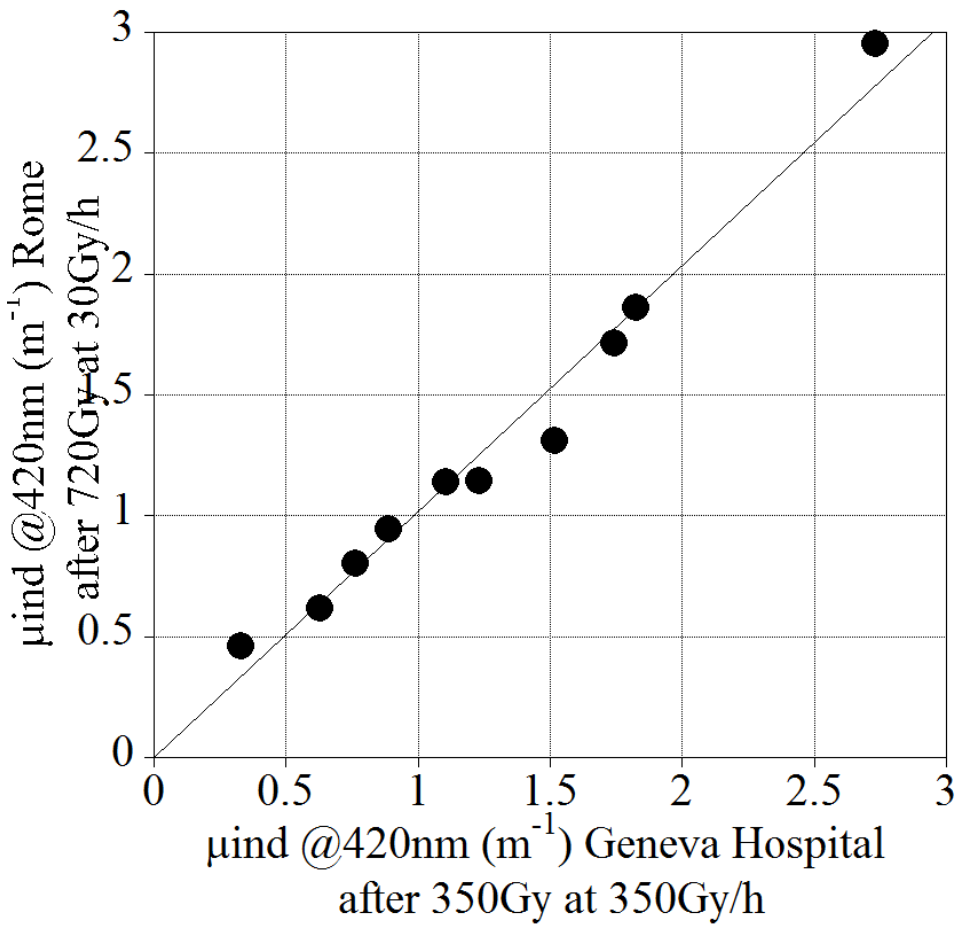

Figure 7.1.1 Correlation between irradiation tests results performed in CERN and Rome regional centers

However, in the case of large scale production at BTCP it was not practical to check the radiation hardness of each crystal individually. Therefore studies were made during the R&D period to find correlations between radiation hardness behavior and optical parameters that can be measured easily for each crystal [42]. It was found that the presence of an initial absorption in the band edge region (350-360 nm) is correlated with poor radiation hardness [27]. Based on this correlation certification limits were defined for transmission spectra [42] which were systematically measured for each crystal using the ACCOR and ACCOS machines. Transmission runs consisted in measuring the optical transmission along the crystal (beam direction from front face to rear face) and transversely at several points along the crystal length. Crystals having one of these optical certification parameters close to or above acceptance limits were subject to systematic irradiation tests. All other crystals were randomly irradiated in both regional centers either at high dose rate (350 Gy/h in Geneva hospital for the CERN regional centre or 30 Gy/h at the Calliope reactor for the Rome regional centre) or under conditions comparable to those in the LHC (0.15 Gy/h front irradiation with the low dose rate $^{60}$Co source at the CERN regional centre). For high dose rates, the limiting value of 1.5 m$^{-1}$ for the induced absorption at 420 nm was used as the qualification parameter. For low dose rates the light yield loss was measured directly and compared to the specification limit of 6% [42]. Crystals which did not fulfill these conditions were rejected.The reliability of these tests was checked on crystals subject to irradiation tests in both regional centers. Fig. 7.1.1 shows the good correlation between the results of irradiation tests performed in the CERN and Rome regional centers.

The radiation hardness over the entire crystal production period (8 years for BTCP and 2 years for SICCAS crystals) was monitored through quality control procedures established for each of two suppliers as follows:

**BTCP Protocol:** agreed between BTCP, INP-Minsk and CERN at the beginning of crystal production in 1998, was a three step radiation hardness certification protocol:

1. A sampling test of the radiation induced absorption performed at INP-Minsk on the top and bottom parts of ingots, aimed at controlling different technological aspects which influence the nature of crystal defects, the longitudinal distribution of the defects along the ingots, the damage recovery and the afterglow.
2. A sampling test performed at BTCP irradiation facilities on machined crystals. Crystals at BTCP were grown by Czochraski method. Several successive growths (crystallizations) were performed in a same crucible by adding new raw material at the end of each crystallization. A maximum number of 13 successive crystallizations was admitted for ECAL crystals. The selected crystals for radiation test were chosen among crystals coming from different crystallizations. Some crystals obtained from the first crystallization were selected to control the stoechiometry of the raw material, other crystals obtained from the 5$^{th}$ crystallization were chosen to control the doping concentration, and all crystals obtained from 9$^{th}$ to 13$^{th}$ crystallization were tested.
3. A sampling irradiation test under high or low irradiation rate performed at CERN. From each batch received, in addition to the crystals with optical certification parameters close or above acceptance limits mentioned above, a randomly chosen set of crystals was tested. Approximately 3% of the crystals produced at BTCP were subjected to this kind of test.

**SICCAS Protocol:** agreed between SICCAS and CERN was a two step radiation hardness certification protocol:

1. All crystals delivered by SICCAS were subject to an irradiation test at the producer consisting of 700 Gy irradiation exposure at a dose rate of 30 Gy/h. Crystals with a radiation induced absorption coefficient above 1.6 $m^{-1}$ were rejected by the producer.
2. Crystals delivered to CERN were further subject to a sampling test performed in both regional centres (CERN and Rome) in order to check the coherence between induced absorption measured using CMS facilities and SICCAS facilities. In addition all crystals having an induced absorption value measured at SICCAS close to the specification limit were systematically irradiated at the Calliope irradiation plant in Rome.

The higher value of the limit of radiation induced absorption coefficient (1.6 $m^{-1}$) in the case of crystals produced at SICCAS is mainly due to their higher LY which makes the corresponding higher LY loss still acceptable for ECAL energy resolution constraints.

## 7.2. Certification of crystal mass production

Crystal production was completed in March 2007 for the Barrel and in March 2008 for the Endcaps. A total of 61335 Barrel crystals were produced in BTCP from September 1998 to March 2007 and 1825 at SICCAS from June 2005 to April 2007. Starting in March 2007, 12015 and 2668 Endcap crystals respectively were produced by the two suppliers.

### 7.2.1 Testing radiation damage at full saturation

Typical induced absorption spectra obtained for BTCP crystals and SICCAS crystals after high dose rate irradiation have already been shown in Fig. 5.1.1. Fig. 7.2.1 shows the distribution of induced absorption measured at 420 nm after irradiation in a Geneva Cantonal Hospital (at a high dose rate of 350 Gy/h for an integrated dose of 350Gy) for a random sampling of BTCP Barrel and Endcap crystals having optical characteristics within ECAL specifications.

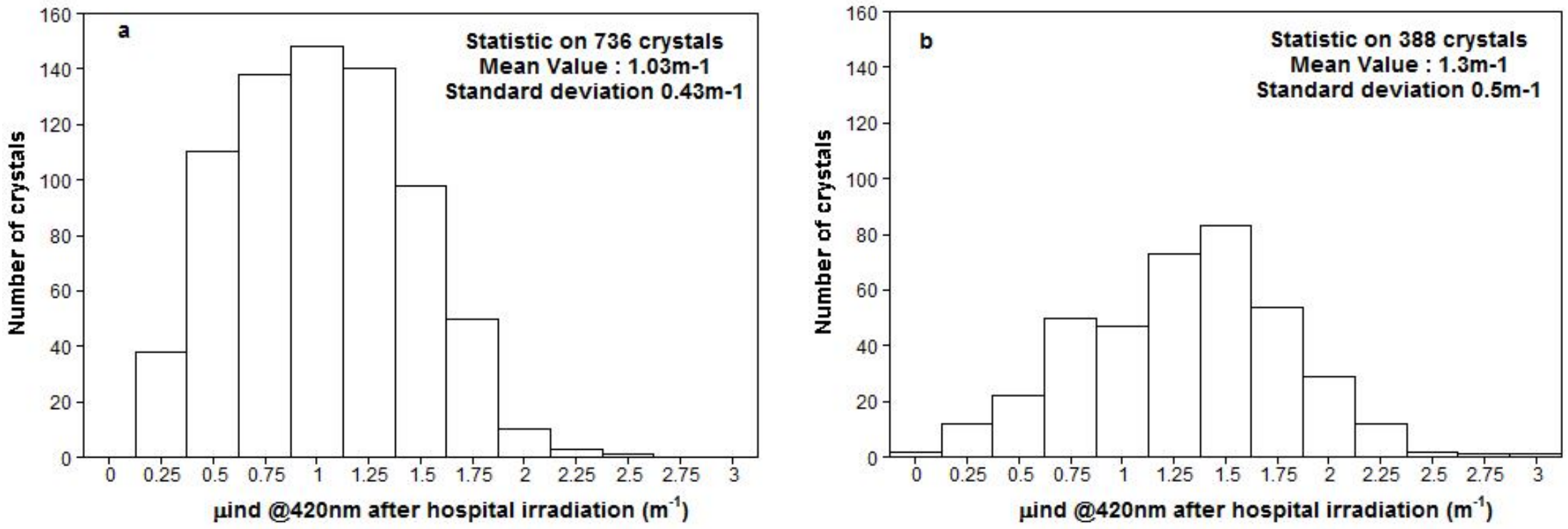

Fig 7.2.1 Distribution of induced absorption at 420 nm after 350 Gy at a dose rate of 350 Gy/h for: (a) BTCP Barrel crystals b) BTCP Endcaps crystals, randomly selected among those having optical parameters within ECAL certification limits

Fig. 7.2.2 shows the correlation between the induced absorption obtained in the ECAL regional centers and at SICCAS for the Barrel and the Endcap crystals. For both producers, for all the crystals tested, an average value of about 1$m^{-1}$ for the induced absorption at full saturation of the damage was obtained.

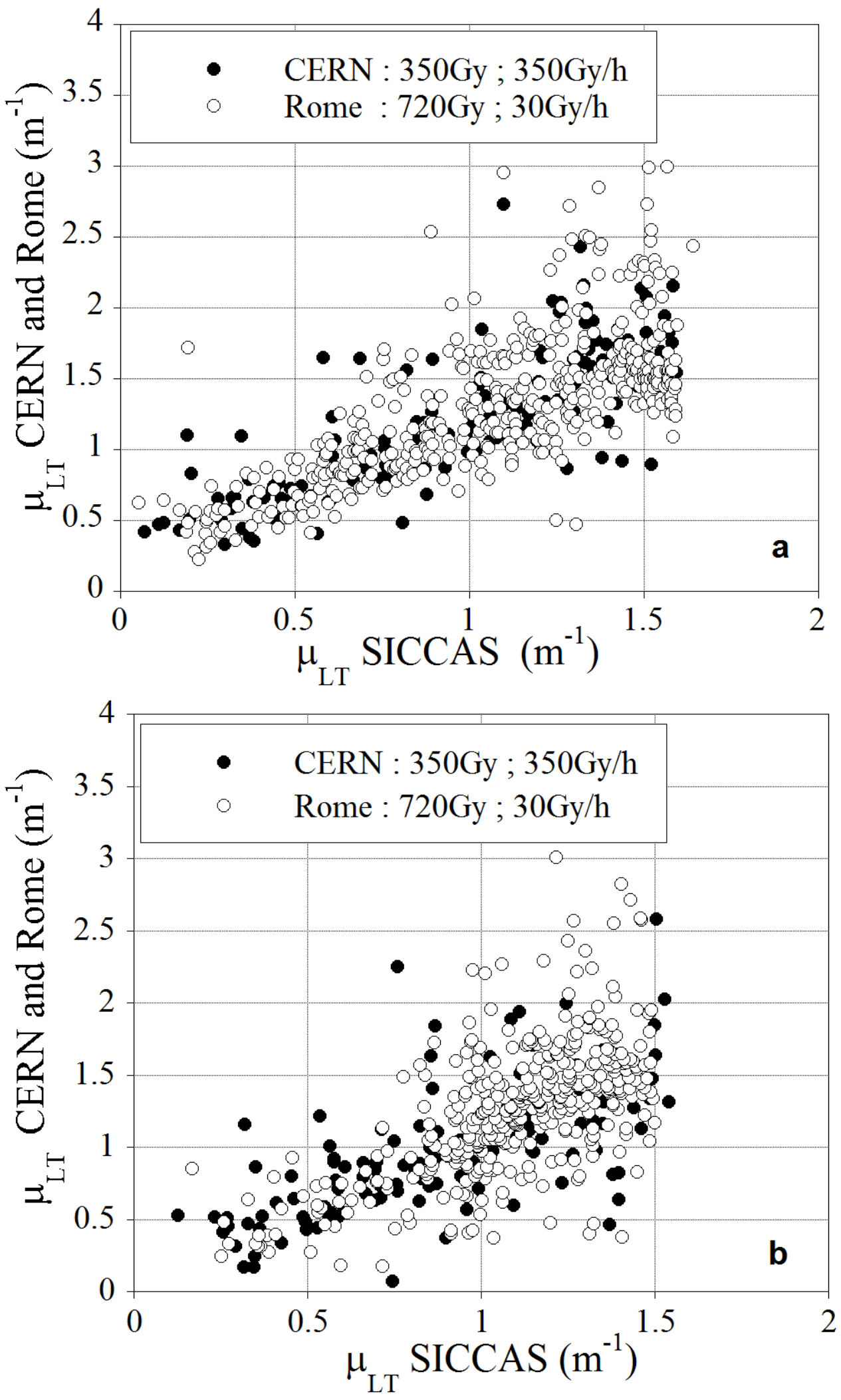

Fig 7.2.2 Correlation between induced absorption measured at SICCAS (700 Gy, 30 Gy/h) and in ECAL regional centres. (a) Barrel crystals, (b) Endcaps crystals.

### 7.2.2 Testing radiation damage in LHC-like conditions

Fig. 7.2.3 shows the distribution of LY loss under front irradiation at LHC radiation levels (0.15Gy/h), for a random sampling of BTCP Barrel crystals. The average light yield loss is 2.4%. Fig. 7.2.4 shows the same for a random sampling of SICCAS Barrel crystals. The average light yield loss is 1.6%.

The radiation damage level observed after irradiation in the LHC like conditions for the ECAL at both producers is below the specification limit of 6% and will guarantee the ability to monitor precisely the crystals transparency.

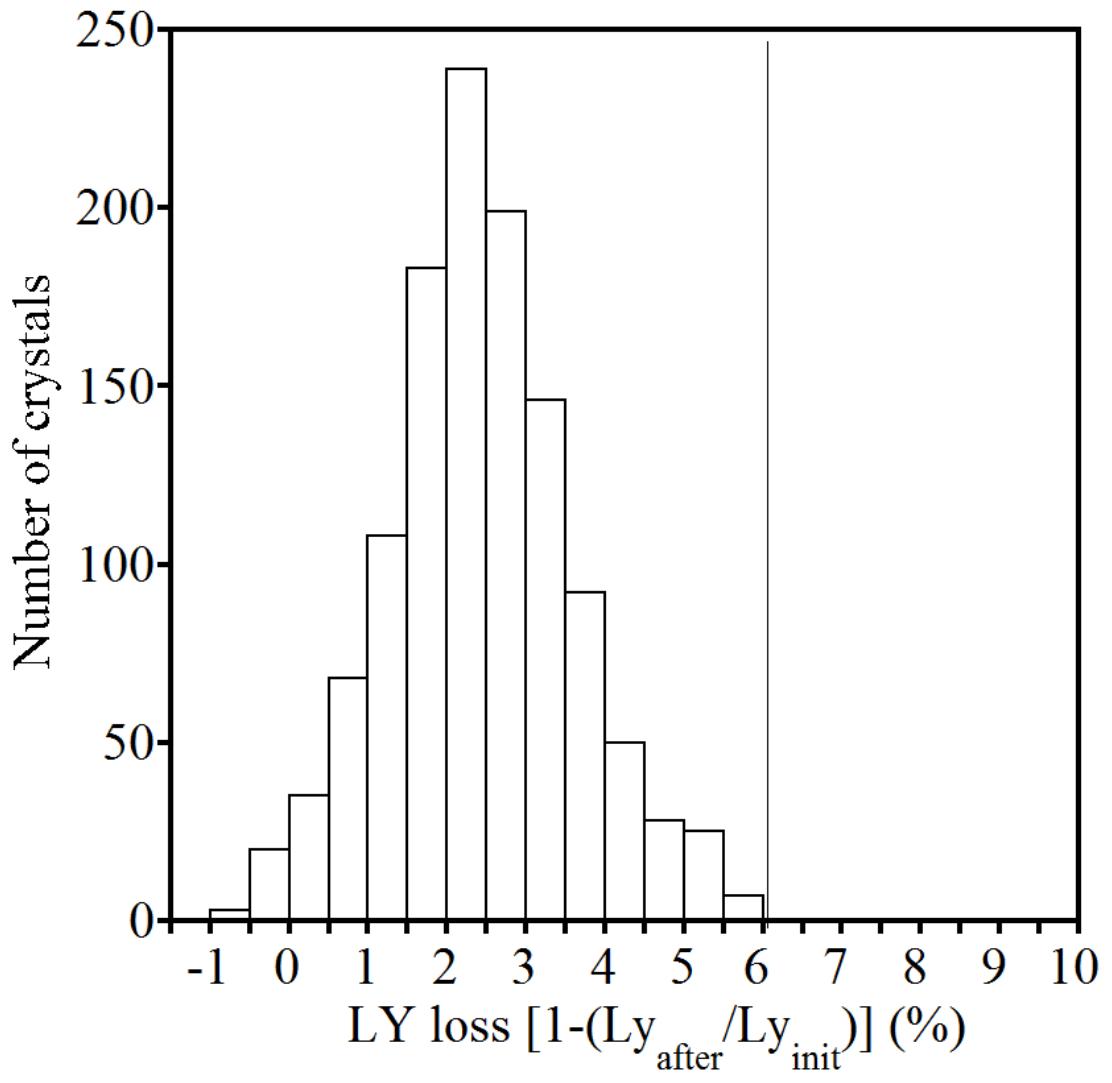

Fig 7.2.3 Distribution of relative LY loss after 1.5 Gy at a rate 0.15 Gy/h for 1203 BTCP crystals randomly selected among those having optical parameters within ECAL certification limits (vertical line defines the 6% certification limit for light yield loss)

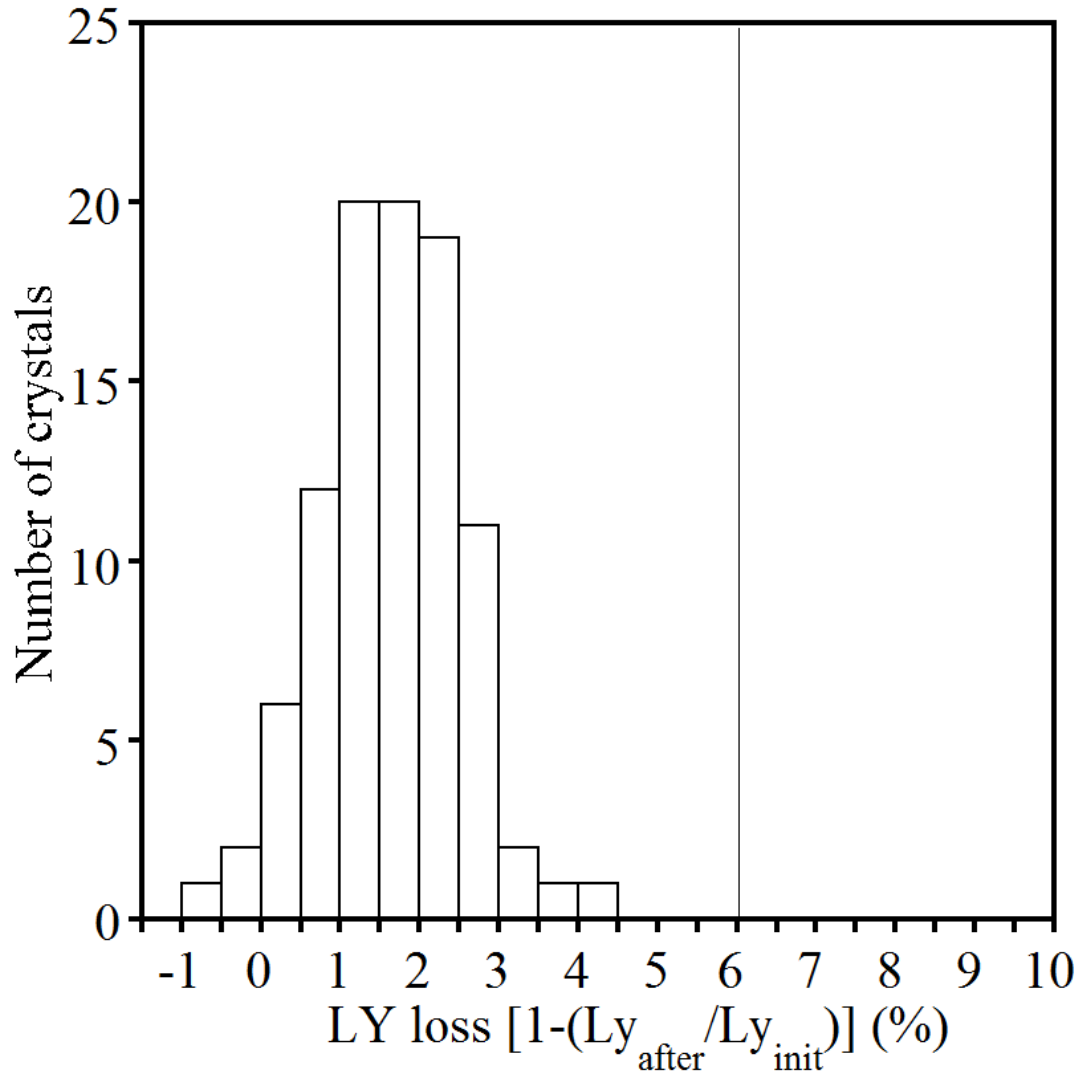

Fig 7.2.4 Distribution of relative LY loss after 1.5 Gy at a rate 0.15 Gy/h for a random sample of 96 SICCAS crystals

## Conclusions

This article has described the investigations that have led to a detailed understanding of $PbWO_4$ scintillation characteristics and radiation induced color centers. These activities were a fundamental contribution to enabling the production of $PbWO_4$ crystals with characteristics satisfying the CMS requirements. The certification of such production extended over several years and quality assurance results have been presented which demonstrate the reliability of the $PbWO_4$ used in CMS.

The excellent linear relationship between the variation of crystal light yield and longitudinal transmission in the wavelength region of 420 - 440 nm in repeated cycles of irradiation-recovery, allows a reliable monitoring of the light yield in LHC exploitation conditions.

## Appendix
## (list of irradiation facilities used in this work)

### A.1 Gamma irradiations

**Geneva Hospital $^{60}$Co source**. Crystals are irradiated from the side with a dose rate of 350 Gy/h. Variations of the crystal transversal and longitudinal transmission, light yield and scintillation kinetics are measured 40 min after the irradiation by a set of spectrometers located at CERN after transport in a thermally isolated box

**CERN – TIS-B27 $^{60}$Co source**. This setup allows the front irradiation of 6 crystals (including one or 2 for reference) at a dose rate of 0.15Gy/h (CMS ECAL Barrel exposure under normal LHC conditions) and an integrated dose of up to 3Gy. The radiation induced light yield is recorded during the irradiation, allowing a direct monitoring of the light transmission loss.

**CERN GIF-X5 beam facilities** including a 137Cs source (lateral irradiation at 0.15Gy/h up to 2 Gy integrated dose) and the possibility to probe the crystal in parallel to the irradiation with a muon or an electron beam.

**CERN H4 beam** Crystals are used in the same configuration as in the CMS experiment, with the same photodetectors, readout electronics, thermal stabilization and monitoring systems. High energy electron beams are shot in the crystals with intensities allowing to mimic the LHC expected dose rate of about 0.15Gy/h.

**COCASE Facility, at CEA/IRFU, Saclay**. It consists of a 60Co source (1.2 MeV, 14 Ci) with geometrical tuning of the dose rate in the region below 1 Gy/h. Relative change of the monitoring signals provided by a Xe pulse lamp and a fast tuning monochromator in the spectral region 380-800 nm is measured. Longitudinal transmission is also measured by a Perkin-Elmer spectrophotometer in the region 300-850 nm before and after the crystal irradiation.

**Institute for Nuclear Problems (INP) $^{60}$Co source facility, Minsk**. It consists of a 60Co well shaped source with fixed dose rate 2300 Gy/h. Variations of the crystal transmission are measured 60 min after irradiation by a Varian "Cary1E" spectrophotometer in the range 300-900 nm. Control of other scintillation parameters is also carried out.

**Institute for Nuclear Research (INR) irradiation facility, Moscow**. A microtron MK-25 and linear electron accelerator (both of 25 MeV electron energy) are used as electromagnetic radiation sources. Dose rates are tuned from 0.06 to 1800 Gy/h. Longitudinal transmission is measured in the region of 390 to 650 nm during the irradiation by a specially developed spectrophotometer based on a transparent grating.

**ENEA-Casaccia, Rome *Calliope* $^{60}$*Co* plant**. This is a 24000Ci (July 2002) source in a water pool. The irradiation plant allows irradiation up to dose rates of 20 kGy/h. Different dosimetric positions are available for $PbWO_4$ radiation hardness study in the range from 10 to 400 Gy/h. Standard $PbWO_4$ tests are performed at irradiation points: 11.8Gy/h, 34.2Gy/h and 350Gy/h (values at October 2006) where PWO crystals are subject to irradiation cycles. For a given dose rate the typical irradiation cycle consists in gamma radiation exposures at increasing doses until the saturation of the induced absorption coefficient is reached. Irradiation is followed by room temperature recovery until acquired data allow for a reliable estimation of the time component(s) of the recovery process. Transmission and LY measurements are performed at typically 15 min after the irradiation exposure is stopped. During the irradiation cycle(s) crystal are kept at room temperature, in the dark in order to avoid recovery processes induced by light.

**CALTECH $^{60}$Co 50Ci source**. Crystals in a light tight package are irradiated from the side with a dose rate ranged from 0.1 to 10 Gy/h by placing samples at different distances to the source.

**Imperial College/Brunel Univeristy $^{60}$Co source**.

**Eichlabor, PSI-Villigen $^{60}$Co source**.

**CALTECH 137Cs 2000Ci source**. Crystals in a light tight package are irradiated from the side with a dose rate of 90 Gy/h when placed at the center of the irradiation chamber with a uniformity of dose rate about 5%. Up to 360 Gy/h may be achieved in the closest position to the source. Variations of the crystal's optical transmission, photo-luminescence, light output and decay kinetics are measured about 5 minutes after the irradiation respectively by a PerkinElmer Lambda-950 spectrometer, Hitachi-F4500 fluorescence spectrophotometer and a Hamamatsu R2059 PMT and LeCroy 3001 QVT charge integrator based pulse height spectrometer. All measurements are carried out in the dark, or under a red light, at 18°C.

In all irradiation facilities mentioned above, a common recovery protocol was defined for PWO crystals of ECAL-CMS. Recovery of radiation damage was obtained by a thermal bleaching process consisting in:

- ramp-up from room temperature to 200°C in not less than 3 hours
- plateau of 6 hours at 200°C
- power stop and natural cooling

### A.2 Neutron irradiations

**Saclay Ulysse reactor**: is a Uranium metal reactor with a fast neutron (about 1MeV) flux ranging from $10^6$ to $10^{11}$ neutrons $cm^{-2}s^{-1}$. Irradiation tests with neutron fluences of $10^{12}$, $2{\cdot}10^{13}$ and $2{\cdot}10^{14}$ neutrons $cm^{-2}$ have been made on several crystals.

**ENEA-Casaccia, Rome *TAPIRO* reactor**: is a fast neutrons reactor particularly adapted for small size experiments due to its relatively low power (5 kW or $4.3 \cdot 10^{14}$ fission neutrons/s) and very small, highly enriched core (11cm height, 12cm diameter, 93.5% enriched Uranium).

### A.3 Charged hadron irradiations

In order to test the possible role of charged hadrons induced lattice defects (stars) some crystals have been exposed to intense pion or proton beams at **PSI-Villigen** and **CERN-PS-IRRAD1** In this last case, crystals were exposed to 20 GeV proton beams with a flux ranging from $10^{12}$ to $10^{13}$ p $cm^{-2}h^{-1}$.

## References

[1] CMS Collaboration, **CERN/LHCC 94–38 (1994)**, *"The Compact Muon Solenoid Technical Proposal"*.

[2] CMS Collaboration, **CERN/LHCC 97–33 (1997)**, *"ECAL Technical design report"*.

[3] M. Huhtinen, P. Lecomte, D. Luckey, F. Nessi-Tedaldi, F. Pauss, **Nuclear Instruments and Methods in Physics Research A 545 (2005) 63–87**, *"High-energy proton induced damage in $PbWO_4$ calorimeter crystals"*.

[4] A.N. Annenkov, A.A. Fedorov, P.-H. Galez, V.A. Kachanov, M.V. Korzhik, V.D. Ligun, J.-M. Moreau, V.N. Nefedov, V.B. Pavlenko, J.-P. Peigneux, T.N. Timoshchenko, B.A. Zadneprovskii, **Physica Status Solidi (a) 156 (1996) 493–504**, *"The influence of additional doping on the spectroscopic and scintillation parameters of $PbWO_4$ crystals"*.

[5] M. Nikl, K. Nitsch, S. Baccaro, A. Cecilia, M. Montecchi, B. Borgia, I. Dafinei, M. Diemoz, M. Martini, E. Rosetta, G. Spinolo, A. Vedda, M. Kobayashi, M. Ishii, Y. Usuki, **Journal of Applied Physics 82 (1997) 5758–5762**, *"Radiation induced formation of color centers in $PbWO_4$ single crystals"*.

[6] E. Auffray, P. Lecoq, M. Korzhik, A. Annenkov, O. Jarolimek, M. Nikl, S. Baccaro, A. Cecilia, M. Diemoz and I. Dafinei, **Nuclear Instruments and Methods in Physics Research A 402 (1998) 75–84**, *"Improvement of several properties of lead tungstate crystals with different doping ions"*.

[7] M. Kobayashi, Y. Usuki, M. Ishii, T. Yazawa, K. Hara, M. Tanaka, M. Nikl, S. Baccaro, A. Cecilia, M. Diemoz and I. Dafinei, **Nuclear Instruments and Methods in Physics Research A 404 (1998) 149–156**, *"Improvement in radiation hardness of $PbWO_4$ scintillating crystals by La-doping"*.

[8] A. Annenkov, E. Auffray, A. Borisevich, M. Korzhik, P. Lecoq, V. Ligun, **Nuclear Instruments and Methods in Physics Research A 426 (1999) 486–490**, *"Suppression of the radiation damage in lead tungstate scintillation crystal"*.

[9] M. Kobayashi, Y. Usuki, M. Ishii, N. Senguttuvan, K. Tanji, M. Chiba, K. Hara, H. Takano, M. Nikl, P. Bohacek, S. Baccaro, A. Cecilia and M. Diemoz, **Nuclear Instruments and Methods in Physics Research A 434 (1999) 412–423**, *"Significant improvement of $PbWO_4$ scintillating crystals by doping with trivalent ions"*.

[10] A.A. Annenkov, M.V. Korzhik, P. Lecoq, **Nuclear Instruments and Methods in Physics Research A 490 (2002) 30–50**, *"Lead tungstate scintillation material"*.

[11] I.Dafinei, E.Auffray, P.Lecoq, M.Schneegans, in **Scintillator and Phosphor Materials, M.J. Weber et al eds, Material Research Society Symposium Proceedings 348 (1994) 99–104**, *"Lead Tungstate for High Energy Calorimetry"*.

[12] S.Baccaro, L.Barone, B.Borgia, F.Castelli, F.Cavallari, I.Dafinei, F.de Notaristefani, M.Diemoz, A.Festinesi, E.Leonardi, E.Longo, M.Montecchi, G.Organtini, **Nuclear Instruments and Methods in Physics Research A 385 (1997) 209–214**, *"Ordinary and extraordinary complex refractive index of the Lead Tungstate ($PbWO_4$) crystal"*.

[13] Peizhi Yang, Jingying Liao, Bingfu Shen, Peifa Shao, Haihong Ni, Zhiwen Yin, **Journal of Crystal Growth 236 (2002) 589–595**, *"Growth of large-size crystal of $PbWO_4$ by vertical Bridgman method with multi-crucibles"*.

[14] J.-M. Moreau, R.E. Gladyshevskii, Ph. Galez, J.-P. Peigneux, M.V. Korzhik, **Journal of Alloys and Compounds 284 (1999) 104–107**, *"A new structural model for Pb-deficient $PbWO_4$"*.

[15] R. Chipaux, G. André, A. Cousson, **Journal of Alloys and Compounds 325 (2001) 91–94**, *"Crystal structure of lead tungstate at 1.4 and 300 K"*.

[16] P. Bohacek, M. Nikl, J. Novak, Z. Malkova, B. Trunda, J. Rysavy, S. Baccaro, A. Cecilia, I. Dafinei, M. Diemoz and K. Jurek, in **Tungstate Crystals, Proceedings of the International Workshop on Tungstate Crystals, S. Baccaro et al eds., Rome, 1998 October 12–14, ISBN 88-87242-10-0, pp 57–60**, *"Stoechiometry and radiation damage of PWO crystals grown from melts of different composition"*.

[17] J. Sykora, M. Husák, O. Jarolímek, A. Strejc, D. Sedmidubsky, **ibidem, pp 95–97**, *"The crystallographical study of compounds originating in the PbO-$WO_3$ system"*.

[18] K. Tanji, M. Ishii, Y. Usuki, M. Kobayashi, K. Hara, H. Takano and N. Senguttuvan, **Journal of Crystal Growth 204 (1999) 505–511**, "*Crystal growth of $PbWO_4$ by the vertical Bridgman method : Effect of crucible thickness and melt composition*".
[19] Baoguo Han, Xiqi Feng, Guanqin Hu, Yanxing Zhang, and Zhiwen Yin, **Journal of Applied Physics 86 (1999) 3571–3575**, *"Annealing effects and radiation damage mechanisms of $PbWO_4$ single crystals"*.
[20] A. Annenkov, V. Ligun, E. Auffray, P. Lecoq, S. Gninenko, N. Golubev, A. Fedorov, M. Korzhik, A. Lobko, O. Missevitch, J.P. Peigneux, Yu.D. Prokoskin, A. Singovski, **CMS Note 1997/008 (1997)**, *"Radiation damage kinetics in PWO crystals"*.
[21] E. Auffray, G. Chevenier, M. Freire, P. Lecoq, J.-M. Le Goff, R. Marcos, G. Drobychev, O. Missévitch, A. Oskine, R. Zouevsky, J.-P. Peigneux, M. Schneegans, **Nuclear Instruments and Methods in Physics Research A 456 (2001) 325–341**, *"Performance of ACCOS, an Automatic Crystal quality Control System for the PWO crystals of the CMS calorimeter"*.
[22] S. Baccaro, B. Borgia, M. Castellani, A. Cecilia, I. Dafinei, M. Diemoz, S. Guerra, E. Longo, M. Montecchi, G. Organtini, F. Pellegrino, **Nuclear Instruments and Methods in Physics Research A 459 (2001) 278–284**, *"An automatic device for the quality control of large-scale crystal's production"*.
[23] M. Montecchi, S. Baccaro, I. Dafinei, M. Diemoz, R.M. Montereali, F. Somma, **The Review of Scientific Instruments 75 (2004) 4636–4640**, "*Lumen: A highly versatile spectrophotometer for measuring the transmittance throughout very long samples as well as microstructures*".
[24] S. Baccaro, L.M. Barone, B. Borgia, F. Castelli, F. Cavallari, F. de Notaristefani, M. Diemoz, R. Faccini, A. Festinesi, E. Leonardi, E. Longo, M. Montecchi, G. Organtini, L. Pacciani, S. Pirro, **Nuclear Instruments and Methods in Physics Research A 385 (1997) 69–73**, *"Precise determination of the light yield of scintillating crystals"*.
[25] I. Dafinei, in **Proc. 8$^{th}$ Int. Conference on Inorganic Scintillators, SCINT2005, A. Getkin and B. Grinyov eds, Alushta, Crimea, Ukraine, 2005 September 19–23, ISBN 9666-02-3884-3, pp 327–330**, *"Optical and scintillation properties of lead tungstate crystals: a statistical approach"*.
[26] M. Nikl, K. Polak, K. Nitsch, E. Mihokova, P. Lecoq, I. Dafinei, P. Reiche, R. Uecker, O. Jarolimek, in **Proc. Int. Conference on Inorganic Scintillators and their Applications, SCINT95, P. Dorenbos and C.W.E. van Eijk eds, Delft, The Netherlands, 1995 August 28 – September 1, ISBN 90-407-1215-8, pp 257–259**, "*Luminescence and scintillation of single $PbWO_4$ crystals*".
[27] E. Auffray, I. Dafinei, F. Gautheron, O. Lafond-Puyet, P. Lecoq, M. Schneegans, **ibidem, pp 282–285**, *"Scintillation characteristics and radiation hardness of PWO scintillators to be used at the CMS electromagnetic calorimeter at CERN"*.
[28] R.Y. Zhu, D.A. Ma, H.B. Newman, C.L. Woody, J.A. Kierstead, S.P. Stoll and P.W. Levy, **Nuclear Instruments and Methods in Physics Research A 376 (1996) 319–334**, *"A study on the properties of lead tungstate crystals"*.
[29] Xiangdong Qu, Liyuan Zhanga, Ren-Yuan Zhu, Jingying Liao, Dingzhong Shen and Zhiwen Yin, **Nuclear Instruments and Methods in Physics Research A 480 (2002) 470–487**, *"A study on yttrium doping in lead tungstate crystals"*.
[30] L.M. Bollinger, G.E. Thomas, **The Review of Scientific Instruments 32(9) (1961) 1044–1050**, *"Measurement of the Time Dependence of Scintillation Intensity by a Delayed-Coincidence Method"*.
[31] S. Baccaro, P. Bohacek, B. Borgia, A. Cecilia, I. Dafinei, M. Diemoz, M. Ishii, O. Jarolimek, M. Kobayashi, M. Martini, M. Montecchi, M. Nikl, K. Nitsch, Y. Usuki, A. Vedda, **Physica Status Solidi (a) 160 (1997) R5–R6**, *"Influence of $La^{3+}$-Doping on Radiation Hardness and Thermoluminescence Characteristics of $PbWO_4$"*.
[32] S. Baccaro, P. Bohacek, B. Borgia, A. Cecilia, S. Croci, I. Dafinei, M. Diemoz, P. Fabeni, M. Ishii, M. Kobayashi, M. Martini, M. Montecchi, M. Nikl, K. Nitsch, G. Organtini, G.P. Pazzi, Y. Usuki, A. Vedda, **Physica Status Solidi (a) 164 (1997) R9–R10,** *"Radiation Damage and Thermoluminescence of Gd-Doped $PbWO_4$"*.
[33] M. Kobayashi, Y. Usuki, M. Ishii, T. Yazawa, K. Hara, M. Tanaka, M. Nikl and K. Nitsch, **Nuclear Instruments and Methods in Physics Research A 399 (1997) 261–268,** *"Improvement in transmittance and decay time of $PbWO_4$ scintillating crystals by La-doping"*.
[34] A.N. Annenkov, E. Auffray, R. Chipaux, G.Yu. Drobychev, A.A. Fedorov, M. Geleoc, N.A. Golubev, M.V. Korzhik, P. Lecoq, A.A. Lednev, A.B. Ligun, O.V. Missevitch, V.B. Pavlenko, J.-P. Peigneux, A.V. Singovski, **Radiation Measurements 29 (1998) 27–38,** *"Systematic Study of the Short-term Instability of $PbWO_4$ Scintillator Parameters under Irradiation"*.
[35] A. Annekov, E. Auffray, M. Korzhik, P. Lecoq, J.-P. Peigneux, **Physica Status Solidi (a) 170 (1998) 47–62**, *"On the Origin of the Transmission Damage in Lead Tungstate Crystals under Irradiation"*.

[36] M.V. Korzhik, V.B. Pavlenko, T.N. Timoschenko, V.A. Katchanov, A.V. Singovskii, A.N. Annenkov, V.A. Ligun, I.M. Solskii, J.-P. Peigneux, **Physica Status Solidi (a) 154 (1996) 779–788**, *"Spectroscopy and Origin of Radiation Centers and Scintillation in $PbWO_4$ Single Crystals"*.
[37] A.N. Annenkov et al., **Physica Status Solidi (a) 191 (2002) 277–290**, *"On the Mechanism of Radiation Damage of Optical Transmission in Lead Tungstate Crystal"*.
[38] Q. Deng, Z. Yin, R.Y. Zhu, **Nuclear Instruments and Methods in Physics Research A 438 (1999) 415–420**, *"Radiation-induced color centers in La-doped $PbWO_4$ crystals"*.
[39] Kazuhiko Hara, Mitsuru Ishii, Masaaki Kobayashi, Martin Nikl, Hideaki Takano, Masashi Tanaka, Kiyokazu Tanji and Yoshiyuki Usuki, **Nuclear Instruments and Methods in Physics Research A 414 (1998) 325–331,** *"La-doped $PbWO_4$ scintillating crystals grown in large ingots"*.
[40] I. Dafinei et al., in **Proc. Int. Conf. on Inorganic Scintillators and Their Applications, SCINT97, Y. Zhiwen et al. eds, 1997 September 22-25, Shanghai, China, pp 219–222**, "*Colour centres production in $PbWO_4$ crystals by UV light exposure"*.
[41] E. Auffray, E. Baguer Battalla, P. Lecoq, S. Paoletti, M. Schneegans, **ibidem pp 199–202**, *"Progress in the radiation hardness of PWO scintillators for CMS calorimeter"*.
[42] E. Auffray, M. Lebeau, P. Lecoq, **CMS NOTE 1998/038 (1998)**, "*Specifications for Lead Tungstate Crystals Preproduction"*.
[43] E. Auffray, **IEEE Transactions on Nuclear Science 55 (2008) 1314–1320**, *"Overview of the 63000 PWO barrel crystals for CMS_ECAL production"*.
[44] Jianming Chen, Rihua Mao, Liyuan Zhang, Ren-Yuan Zhu, in **Nuclear Science Symposium Conference Record, 2005 IEEE, Volume 1, 2005, October 23–29, pp 249–253**, *"A damage and recovery study for lead tungstate crystal samples from BTCP and SIC"*.
[45] P. Lecoq, A. Annenkov, A. Gektin, M. Korzhik, C. Pedrini, ***"Inorganic Scintillators for Detector Systems: Physical Principles and Crystal Engineering"***, ISSN 1611-1052, **Springer, Berlin, (2006)**.
[46] C. D'Ambrosio, C. Ercoli, S. Jaaskelainen, E. Rosso, P. Wicht, **Nuclear Instruments and Methods in Physics Research A 388 (1997) 119–126**, *"Low dose-rate irradiation set-up for scintillating crystals"*.
[47] M.Kh. Ashurov, Sh.Kh. Ismoilov, K. Khatamov, É.M. Gasanov, I.R. Rustamov and V.A. Kozlov, **Atomic Energy 91 (2001) 560–563**, *"Radiation-Induced Optical Absorption in $PbWO_4$:La Scintillation Crystals"*.
[48] R. Chipaux, J.-L. Faure, J.-P. Pansart, J. Poinsignon, P. Rebourgeard, J. Tartas, G. Dauphin, J. Safieh, in **Scintillator and Phosphor Materials, M.J. Weber et al eds, Material Research Society Symposium Proceedings 348 (1994) 481–486**, *"Study of neutron damage resistance of some scintillating crystals and associated photodetectors with the nuclear reactor ULYSSE"*.
[49] R. Chipaux, O. Toson, in **Proc. Int. Conference on Inorganic Scintillators and their Applications, SCINT95, P. Dorenbos and C.W.E. van Eijk eds, Delft, The Netherlands, 1995 August 28 – September 1, ISBN 90-407-1215-8, pp 274–277**, also **CMS/TN 95–126** ,*"Resistance of lead tungstate and cerium fluoride to low rate gamma irradiation or fast neutrons exposure"*.
[50] R. Chipaux, M.V. Korzhik, A. Borisevich, P. Lecoq, C. Dujardin, in **Proc. 8th Int. Conference on Inorganic Scintillators, SCINT2005, A. Getkin and B. Grinyov eds, Alushta, Crimea, Ukraine, 2005 September 19–23, ISBN 9666-02-3884-3, pp 369–371**,*"Behaviour of PWO scintillators after high fluence neutron irradiation"*.
[51] P. Lecomte, D. Luckey, F. Nessi-Tedaldi, F. Pauss, **Nuclear Instruments and Methods in Physics Research Section A 564 2006 164–168**, *"High-Energy Proton Induced Damage Study of Scintillation Light Output from $PbWO_4$ Calorimeter Crystals"*.
[52] P. Lecomte, D. Luckey, F. Nessi-Tedaldi, F. Pauss and D. Renker, **Nuclear Instruments and Methods in Physics Research Section A 587 (2008) 266–271,** *"Comparison between high-energy proton and charged pion induced damage in $PbWO_4$ calorimeter crystals"*.
[53] P. Adzic et al**, The European Physical Journal C, Particles and Fields 44 suppl. 2 (2006) 1–10,** *"Results of the first performance tests of theCMS electromagnetic calorimeter"* .
[54] M. Anfreville et al, **Nuclear Instruments and Methods in Physics Research A 594 (2008) 292–320**, *"Laser monitoring system for the CMS lead tungstate crystal calorimeter"*.
[55] Liyuan Zhang, Kejun Zhu, David Bailleux, Adolf Bornheim, and Ren-Yuan Zhu**, IEEE Transactions in Nuclear Science 55 (2008) 637–643**, *"Implementation of a Software Feedback Control for the CMS Monitoring Lasers"*.
[56] Liyuan Y. Zhang, Kejun J. Zhu, Ren-Yuan Zhu, and Duncan T. Liu, **IEEE Transactions in Nuclear Science 48 (2001) 372–378,** *"Monitoring light source for CMS lead tungstate crystal calorimeter at LHC"*.
[57] Liyuan Zhang, David Bailleux, Adolf Bornheim, Kejun Zhu, and Ren-Yuan Zhu, **IEEE Transactions in Nuclear Science 52 (2005) 1123–1130,** *"Performance of the Monitoring Light Source for the CMS Lead Tungstate Crystal Calorimeter"*.